# Foamability of aqueous solutions:
# Role of surfactant type and concentration

B. Petkova,[1] S. Tcholakova,[1*] M. Chenkova,[1]
K. Golemanov,[1] N. Denkov[1], D. Thorley[2] and S. Stoyanov [3,4,5]

[1]*Department of Chemical and Pharmaceutical Engineering*
*Faculty of Chemistry and Pharmacy, Sofia University*
*1 J. Bourchier Ave., 1164 Sofia, Bulgaria*

[2]*Unilever R&D, Port Sunlight, Quarry Road East, Bebington, Wirral CH63 3JW, UK*

[3]*Unilever R&D, Vlaardingen, The Netherlands*

[4]*Laboratory of Physical Chemistry and Colloid Science, Wageningen University, 6703 HB Wageningen, The Netherlands*

[5]*Department of Mechanical Engineering, University College London, Torrington Place, London WC1E 7JE, UK*

*Corresponding author:
Prof. Slavka Tcholakova
Department of Chemical and Pharmaceutical Engineering
Faculty of Chemistry and Pharmacy, Sofia University
1 James Bourchier Ave., 1164 Sofia
Bulgaria

Phone: (+359-2) 962 5310
Fax:     (+359-2) 962 5643
E-mail: SC@LCPE.UNI-SOFIA.BG

**Abstract**

In this paper we study the main surface characteristics which control the foamability of solutions of various surfactants. Systematic series of experiments with anionic, cationic and nonionic surfactants with different head groups and chain lengths are performed in a wide concentration range, from 0.001 mM to 100 mM. The electrolyte (NaCl) concentration is also varied from 0 up to 100 mM. For all surfactants studied, three regions in the dependence of the foamability, $V_A$, on the logarithm of surfactant concentration, lg$C_S$, are observed. In Region 1, $V_A$ is very low and depends weakly on $C_S$. In Region 2, $V_A$ increases steeply with $C_S$. In Region 3, $V_A$ reaches a plateau. To analyse these results, the dynamic and equilibrium surface tensions of the foamed solutions are measured. A key new element in our interpretation of the foaming data is that we use the surface tension measurements to determine the dependence of the main surface properties (surfactant adsorption, surface coverage and surface elasticity) on the surface age of the bubbles. In this way we interpret the results from the foaming tests by considering the properties of the dynamic adsorption layers, formed during foaming. The performed analysis reveals a large qualitative difference between the nonionic and ionic surfactants with respect to their foaming profiles. The data for the nonionic and ionic surfactants merge around two master curves when plotted as a function of the surface coverage, the surface mobility factor, or the Gibbs elasticity of the dynamic adsorption layers. This difference between the ionic and nonionic surfactants is explained with the important contribution of the electrostatic repulsion between the foam film surfaces for the ionic surfactants which stabilizes the dynamic foam films even at moderate surface coverage and at relatively high ionic strength (up to 100 mM). In contrast, the films formed from solutions of nonionic surfactants are stabilized via steric repulsion which becomes sufficiently high to prevent bubble coalescence only at rather high surface coverage (> 90 %) which corresponds to related high Gibbs elasticity (> 150 mN/m) and low surface mobility of the dynamic adsorption layers. Mechanistic explanations of all observed trends are provided and some important similarities and differences with the process of emulsification are outlined.

# Contents

# 1. Introduction.

Surfactants are essential ingredients in laundry, household and personal care products, and in various technological processes. In many of these systems, the foamability of the surfactant solutions appears as desired or undesired phenomenon, depending on the specific application. Therefore, understanding the process and revealing the key physicochemical and hydrodynamic factors which control the foaming process is very important from both scientific and practical viewpoints.

To build a general and universal interpretation of the experimental results about the foamability of surfactant solutions, one should consider the processes of air entrapment and bubble coalescence which have opposite effects on foam volume - see Figure 1. The foam volume increases when a newly entrapped air during mechanical agitation or gas incorporation (via bubbling or from chemical reaction) is unable to coalesce with the large air-water interface. On the opposite, the coalescence between entrapped air bubbles and this large interface removes the trapped air and keeps the foam volume low. On its turn, the bubble coalescence depends on the competition between the rate of surfactant adsorption on the bubble surfaces and the drainage time of the foam films, formed between the air bubbles and the large air-water interface. If the adsorption rate is faster, the coalescence may be suppressed, due to the repulsion between the bubble and the large gas-liquid interface which may arise only when the gas-liquid intefaces are covered with a sufficient amount of adsorbed molecules. In contrast, if the rate of adsorption is slower, the formed foam films rapidly thin to their critical thickness at which the attractive forces between the film surfaces dominate, the foam films break and the bubbles coalesce before the protective adsorption layer is formed.

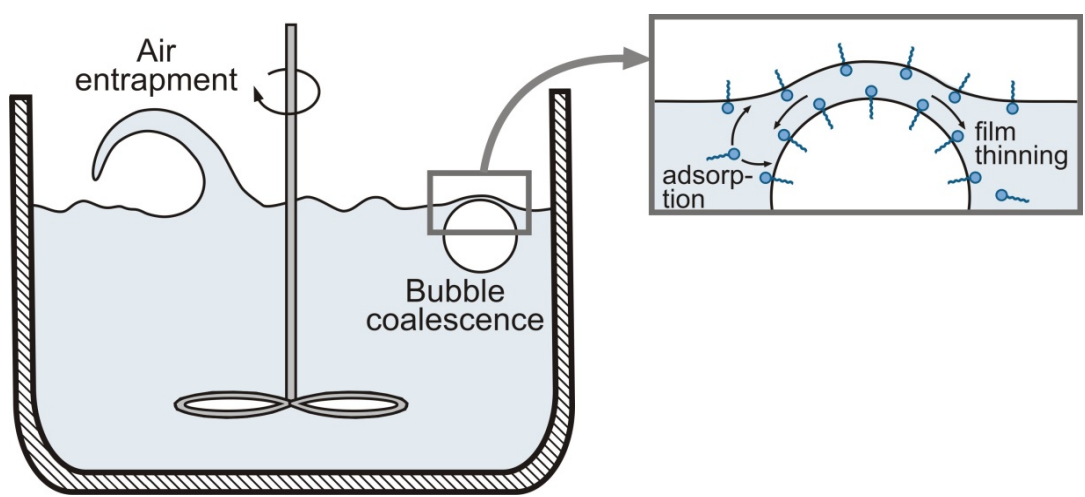

**Figure 1.** Schematic presentation of the main physicochemical processes which define the foam volume upon foaming. The foam volume is determined by the interplay between the processes of air entrapment and bubble coalescence with the large air-water interface. On its turn, the coalescence depends on the competition between the rates of surfactant adsorption and foam film thinning to the critical thickness of film rupture.

The physicochemical analysis of the above concepts is complicated by the fact that the various surfactants may have different stabilizing efficiency at the same surface coverage. For example, one may expect a significant difference between the ionic and nonionic surfactants because the surface forces between the foam film surfaces (electrostatic, steric) are expected to play a crucial role in foam film stabilization. Further complication is that one should consider the surfactant adsorption, surface properties (such as surface coverage and Gibbs elasticity) and surface forces (disjoining pressure) of the dynamic adsorption layers formed during foaming, which usually are very far away from the equilibrium ones.

All these complications lead to the fact that there is no unifying and self-consistent theoretical approach to include the above elements and to describe the available results from the foaming tests. There are different theoretical models which capture the role of one or another factor for foam film rupture and bubble coalescence, but they are all developed for more idealized dynamics of film thinning, e.g. for films with constant diameter and fixed capillary pressure like those formed in a capillary cell or between a large air-water interface and a rising bubble, pushed by buoyancy [1-10]. The relation between the results from such model studies and the results from actual foaming experiments has never been clarified convincingly, mainly due to the enormous complexity of the dynamic processes of foaming.

In the various studies of foaming [11-35] several physicochemical parameters were proposed to explain the variations in the foamability of the surfactants solutions: surfactant concentration [11] and its relation to the critical micellar concentration (CMC); dynamic surface tension (DST) [12-27]; surface mobility expressed through the Marangoni effect [28], surface modulus [29] or surface elasticity [30] of the adsorption layers; stability of the single foam films [31,32] expressed through the disjoining pressure [33] and its components, such as steric repulsion and structural forces [33-35]. Generally speaking, each of these characteristics could be important and their interplay should be understood much better if we want to describe and control the complex process of foam formation.

Most often, it is assumed in the literature that the volume of the generated foam correlates with the rate of surfactant adsorption, which is determined by measuring the DST, and with the amount of adsorbed surfactant at the air-water interface. Many researchers showed in their studies that lower dynamic surface tension often corresponds to higher foaminess of the solutions [12-27]. Also, at concentrations above the CMC, the kinetics of de-micellization and release of surfactant monomers from the micelles was found to play a role [16,17]. Less foam was generated in foaming processes with intensive agitation for surfactants with very low critical micellization concentration (which reduces the equilibrium concentration of monomers in the micellar solutions) and stable micelles, such as those of the nonionic surfactants and with longer chain length.

The observed correlation between the volume of generated foam and DST could be attributed to the more efficient suppression of the bubble coalescence in the case of rapidly adsorbing surfactants. One of the main mechanisms for dynamic stabilization of the freshly formed foam films is the Marangoni effect which may lead to significantly reduced rate of film thinning [15,28]. The Marangoni effect and the related deceleration of film thinning depend strongly on the instantaneous quantity of surfactant adsorbed on the film surface in the moment of film formation. Marangoni effect is related to the surface Gibbs elasticity, $E_G$, which acts to restore the homogeneous distribution of surfactant along the film surface. In [32] the properties of single vertical foam films and the foamability of solutions containing different surfactants were compared. The results showed that only the stability of black films under dynamic conditions has some correlation with the foamability of the same surfactant solutions for different surfactant types.

In other studies it was shown that the changes in the values of the surface dilational modulus may exhibit similar trends to the foamability of the respective solutions for surfactants with different molecular structures [30]. This relation shows that the surface elasticity could be an important factor for the processes of foam formation and stabilization. Also, the bubble break-up is an intrinsic process of foam generation. In a previous article [29] we studied the factors controlling the kinetics of bubble break-up in sheared foams and found that high surface modulus of the surfactant solutions (above 100 mN/m) leads to the formation of much smaller bubbles due to a rapid breakup of the initial bigger bubbles. Furthermore, in a later study [36] we showed that the higher viscoelasticity of the foam containing smaller bubbles may reduce the volume of the formed foam, thus suppressing the solution foamability. Thus we see that the effect of surface elasticity needs further clarification.

Some authors reported an important relation between the foam and surface properties, on one side, and the surfactant molecular structure, on the other side. At concentrations below and above the CMC, the dynamic surface activity was shown to increase with the increase of the molecular mass of the surfactant molecules, while the foamability was found to decrease due to slower diffusion of the surfactant molecules [11]. The length of the hydrophobic tail is identified as a parameter controlling the rate of diffusion, adsorption and arrangement at the interface [18,20-22]. For a given alkyl chain length, increasing the hydrophilicity of the molecules leads to boost in foamability [34]. For foams produced from solutions of small amphiphilic single- and double-tail surfactants, the number of the hydrophobic tails and their length play a crucial role for the foaming while the head group was reported to be of secondary importance [35]. The authors suggested that the critical aggregation concentration could be used as a predictor for the ability of the small amphiphilic molecules to enhance foaming.

All these results indicate that one should analyse much deeper the properties of the dynamic adsorption layers, formed on the bubble surface during foaming, in order to explain the observed trends in the foaming experiments and to identify the key physicochemical factors controlling this process.

Based on the above brief literature overview, we defined the following major aims of the current study:

(1) To study systematically the role of the various physicochemical factors on the foamability of surfactant solutions using a series of seven surfactants which differ in their type (ionic and nonionic), chain length (12 and 16), head group structure and charge (non-ionic Brij and Tween, cationic and anionic) and concentration (up to 100 mM). The role of ionic strength was also studied by varying the concentration of a neutral electrolyte (NaCl) between 0 and 100 mM.

(2) To analyse the experimental data by considering the properties of the dynamic adsorption layers, taking into account their rapid change with the time of surface aging. On this basis, to reveal the key physicochemical characteristics of the adsorption layers which govern the initial rate of foam generation and the volume of accumulated foam for the various surfactants. The idea is to identify those "universal" parameter(s) which could explain the data for the various surfactant solutions studied.

To achieve the above aims, we combine several experimental methods to obtain complementary information about the surface and foaming properties of the various surfactant solutions – foam tests, dynamic and equilibrium surface tension measurements. Self-consistent interpretation of the results obtained by all these methods is proposed. Note that we use a foaming method with very intensive mechanical agitation (Bartsch shaking test) and that many of the studied solutions are of relatively low surfactant concentration (below and around the CMC) – as a result, the bubble coalescence plays a crucial role for the volume of the foams studied. From this viewpoint, the current paper extends and complements our previous study [36] in which only the range of high surfactant concentration was investigated and the bubble coalescence was completely suppressed.

The article is organized as follows: Section 2 describes the materials and methods used. The experimental results are described in Section 3. Their interpretation and discussion is presented in Section 4. The main conclusions are summarized in Section 5.

## 2. Materials and methods.
### 2.1. Materials

The following surfactants are studied: one anionic – sodium dodecyl sulphate (SDS), two cationic – dodecyltrimethylammonium bromide (DTAB) and cetyltrimethylammonium bromide (CTAB); and four nonionic – polyoxyethylene-23 lauryl ether (Brij 35); polyoxyethylene-20 cetyl ether (Brij 58); polyoxyethylenesorbitan monolaurate (Tween 20); polyoxyethylene sorbitan monopalmitate (Tween 40). SDS is product of Acros while all other surfactants were purchased from Sigma. These surfactants have hydrophobic chain of either 12 carbon atoms (SDS, DTAB, Brij 35, Tween 20) or 16 carbon atoms (CTAB, Brij 58, Tween 40).

On purpose, all surfactants were used as received to reproduce the real technical surfactant mixtures which are typically faced in the practical applications. The presence of different components in these technical surfactant samples (e.g. of dodecanol in the SDS sample) is explicitly considered in the analysis of the experimental data.

The aqueous solutions were prepared with deionized water purified by Elix 3 purification system (Millipore, USA). To vary the ionic strength we used NaCl with purity 99.8% (product of Teokom, Bulgaria).

### 2.2. Measurements of the equilibrium and dynamic surface tension of the surfactant solutions.

The equilibrium surface tension of the foaming solutions, $\sigma$, was measured with the Wilhelmy plate method on tensiometer K100 (Kruss GmbH, Germany) at $T = 20$ °C. The dynamic surface tension of the solutions was measured with the maximum bubble pressure method on tensiometer BP2 (Kruss GmbH, Germany) at 20 °C.

### 2.3. Foamability of studied solutions.

We characterized the foamability of the studied solutions using a custom-made, automated Bartsch test (shaken cylinder). The apparatus allows shaking of a 130 mL glass cylinder, which is fixed to a holder. The holder is rotating, so that the axis of the measuring cylinder changes its angle with respect to the vertical: from 0º in the initial position, via 90º (horizontal cylinder), up to 135º, and back. Because the inclination of the cylinder axis continuously changes during the experiment, the solution moves inside the cylinder. The foam is produced mostly in the moments when the solution hits the top and bottom ends of the cylinder, when the cylinder changes its direction of motion. The frequency of the cylinder cyclic motion and the number of cycles are defined via the control panel in the beginning of each experiment. In our experiments, the shaking period was 1.23 s (frequency = 0.813 $s^{-1}$), the volume of the surfactant solution was 10 mL. We determine the amount of the trapped air within 2-5 s after stopping the cylinder agitation to exclude the effect of the possible subsequent collapse of the formed foam at low surfactant concentrations. Due to the specific dynamics of the foaming test used, in which the cylinders hit an obstacle at the end of each shake cycle, the foam is always collected on top of the solution surface, viz. we have no contribution of the undesired "lacing effect" in our measurements. If very big single transient bubble was generated in the cylinder, its volume was excluded from the measured foam volume, because such single bubbles are not integral part of the foam. For each surfactant concentration, at least 3 measurements were performed. For most low-surfactant concentrations, the number of experiments was > 5 to ensure statistically robust results.

The foamability of the studied systems was characterized via the volume of air, $V_A$, trapped in the solution. $V_A$ was calculated by subtracting the volume of the solution (10 mL) from the total volume (solution + foam) measured after a given number of shake cycles. Note that the value of the measured quantity, $V_A$, is not affected by the water drainage from the foam, because only the upper level of the foam is used to determine it. We measured $V_A$ in foams generated after 3, 5, 10, 20, 30, 50 and 100 consecutive shake cycles. All experiments were performed at $T = 20\ °C$.

## 3. Experimental results.

### 3.1. Surface tension isotherms.

To determine the critical micellar concentration, surfactant adsorption at CMC, and the maximal adsorption, we measured the surface tension as a function of time (up to 900 s) of surfactant solutions with concentration varied between $10^{-3}$ mM and 50 mM using the Wilhelmy plate method. The values of $\sigma(t)$ measured between 750 and 900 s were used to construct the dependence $\sigma(t^{-1/2})$ and to determine the equilibrium surface tension at given surfactant concentration from the intercept of the linear dependence in this plot at $t \to \infty$.

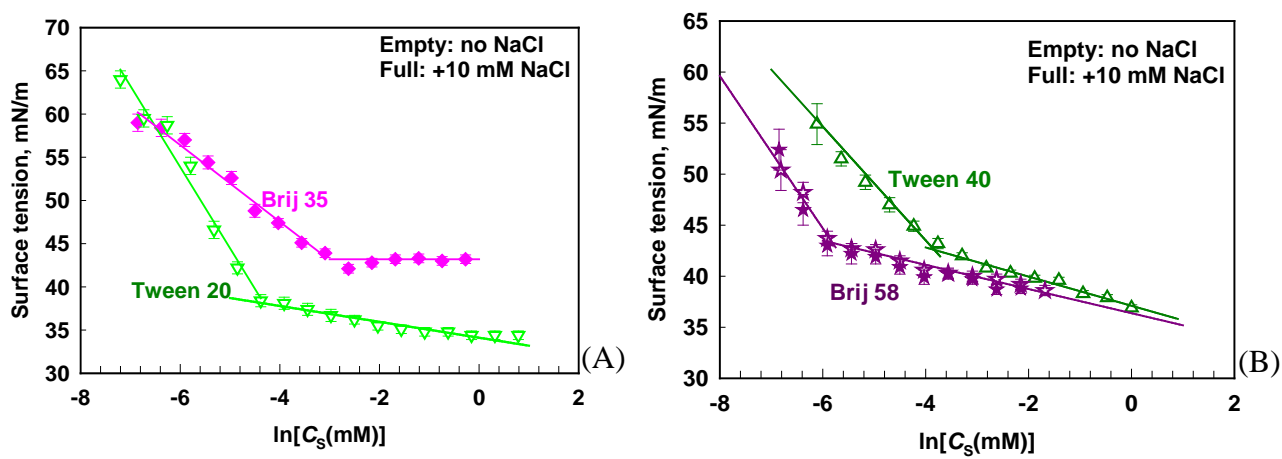

**Figure 2.** Surface tension isotherms for the studied nonionic surfactants: (A) Brij 35 and Tween 20; (B) Tween 40 and Brij 58. In all graphs in the paper, the empty symbols represent data obtained without any additional electrolyte, the full symbols present data obtained in the presence of 10 NaCl, and crossed symbols present data obtained at 100 mM NaCl.

From the data presented in Figure 2 one sees that Brij 35 has the typical behaviour of single surfactant without noticeable contribution of other surface active additives (admixtures), whereas the other nonionic surfactants exhibit a continuous decrease of the surface tension even above the CMC. Similar continuous decrease of the surface tension above the CMC was reported before for Tween 20 adsorption on oil-water interface [37,38]; for nonionic surfactants it is related to a gradual change in the composition of the adsorption layers. These surfactants are technical mixtures of different components which vary in their chain length and in the number of ethoxy groups. Therefore, the composition of their micelles and adsorption layers may vary in the concentration range around and above the CMC [39-41].

As seen from Figure 2, the presence of ad-mixtures in most of the studied surfactants has significant impact on the properties of the adsorption isotherms and we can expect that these ad-mixtures will affect also the foaming properties of these solutions. To gain information about the properties of the formed mixed adsorption layers, we interpreted the measured surface tension isotherms in the following way:

(1) Using the approach of Rehfeld [42] we fit the experimental data for the surface tension vs. surfactant concentration around the CMC by a linear dependance of $\ln C_s$:

$$\sigma = z_0 + z_1 \ln C_S \tag{1}$$

where $z_0$ and $z_1$ are numerical coefficients which are determined from the best fit to the experimental data (see the straigth lines in Figure 2);

(2) We use the Gibbs adsorption isotherm to determine the total surfactant adsorption on the solution surface at the CMC;

(3) Using Volmer adsorption isotherm we determine the average excluded area per molecule in the adsorption layer.

The above approach is very appropriate for mixtures of nonionic surfactant components (as in the case of technical nonionic surfactants). Indeed, for multicomponent mixtures, the Gibbs adsorption isotherm at fixed temperature reads [43]:

$$d\sigma = -\sum_{i=1}^{N} \Gamma_i d\mu_i \tag{2}$$

Here $\sigma$ is surface tension, $\Gamma_i$ is adsorption of $i$-th component on the solution surface, and $\mu_i$ is its chemical potential in the bulk solution. Under the assumption that the bulk surfactant solution can be considered as an ideal solution eq. (2) takes the form [43]:

$$d\sigma = -\sum_{i=1}^{N} \Gamma_i RT d\ln C_i \tag{3}$$

where $R$ is universal gas constant, $T$ is temperature, and $C_i$ is surfactant concentration of the $i$-th component in the solution. For most of the surfactants studied, we have no detailed information about the type and concentration of the various surface active species present. On the other hand, we know the total concentration of surfactant dissolved in the solution, $C_S$, which is related to the concentration of each surfactant component in the solution, $C_i$, through its molar fraction in the mixture, $x_i = C_i/C_S$. Thus, eq. (2) could be represented in the form:

$$d\sigma = -\sum_{i=1}^{N}\Gamma_i RTd\ln C_S = -\sum_{i=1}^{N}\Gamma_i RTd\ln(x_i C_S) = -\Gamma_{tot}RTd\ln C_S \tag{4}$$

Note that in the derivation of eq. (4) we used the fact that the molar fraction of the surfactant components in the surfactant mixture, $x_i$, does not change upon increase of the total surfactant concentration, $C_S$. Hence, the differentiation of the molar fractions $dx_i = 0$ and eq. (4) follows as a rigorous corollary of eq. (3), without any additional approximation.

Here $\Gamma_{tot} = \sum_{i=1}^{N}\Gamma_i$ is the sum of all adsorbed species on the solution surface. Using eq. (4) we can determine the total adsorption, $\Gamma_{tot}$, from the available experimental data for $\sigma(C_S)$. The results from the interpretation of the experimental data for the various surfactants by eq. (4) are summarized in Table 1.

The second step in our analysis includes the assumption that we can apply Volmer adsorption isotherm to describe (approximately) the relation between the surface tension and surfactant adsorption. Indeed, it was shown in ref. [44] that Volmer adsorption isotherm can be used to describe the experimental data for a two-component mixture of nonionic surfactants via the relation:

$$\frac{\pi}{k_B T} = \frac{\Gamma_{tot}}{1 - \alpha \Gamma_{tot}} \tag{5}$$

Here $\pi$ is the surface pressure, $\pi = \sigma_0 - \sigma$, where $\sigma_0$ is the surface tension of the aqueous phase without surfactant, $\sigma(C_S)$ is the equilibrium surface tension at a certain surfactant concentration, $\Gamma_{tot}$ is the total adsorption of the various species, and $\alpha$ is an average excluded area per molecule, which for binary mixture was found to be given by the expression [45]:

$$\alpha \equiv \alpha_{11}X_1^2 + 2\alpha_{12}X_1 X_2 + \alpha_{22}X_2^2 \tag{6}$$

Here $X_i$ is the molar fraction of $i$-th component in the adsorption layer ($i = 1$ or 2), $\alpha_{ii}$ is the excluded area per molecule for this component and $\alpha_{12}$ is defined as [45]:

$$\alpha_{12} = \left(\frac{\sqrt{\alpha_{11}} + \sqrt{\alpha_{22}}}{2}\right)^2 \tag{7}$$

One sees from eqs. (5)-(7) that $\alpha$ plays the role of an apparent excluded area per molecule in the mixed adsorption layer of binary solutions.

In our mixtures we have larger number of components and we do not know their molar fractions on the interface. Therefore we used eq. (5) to determine the value of $\alpha$ which is considered below as an effective average area per molecule in the mixed adsorption layer. For this purpose we determine $\Gamma_{tot}$ from the slope of the surface tension isotherm around the CMC, as shown in Figure 2. Then, we determine the value of π at CMC and, finally, we determine the value of α from the measured value of σ at CMC using eq. (5).

The results from the above analysis are shown for the nonionic surfactants in the first four rows in Table 1, along with representative results from literature [41-53]. One sees that our results are in a relatively good agreement with the literature results with respect to all characteristics studied – CMC, surface tension at CMC, surfactant adsorption at CMC, and average excluded area per molecule.

The only exception is the results for Tween 20. The excluded area per molecule for this surfactant is significantly smaller in our experiments, as compared to the values reported in literature. This difference is most probably due to the presence of surfactant components with smaller number of ethoxy groups in our surfactant sample, which are able to adsorb in between the bulky head groups of Tween 20. This explanation is in a good agreement also with the lower value of the CMC, determined in our study. As shown in Ref. [54] such behaviour could be explained with the presence of surface active components which are able to form compact adsorption layers even before the micelle formation in the bulk solution.

Similar series of experiments were performed with the ionic surfactants, see Figure 3.

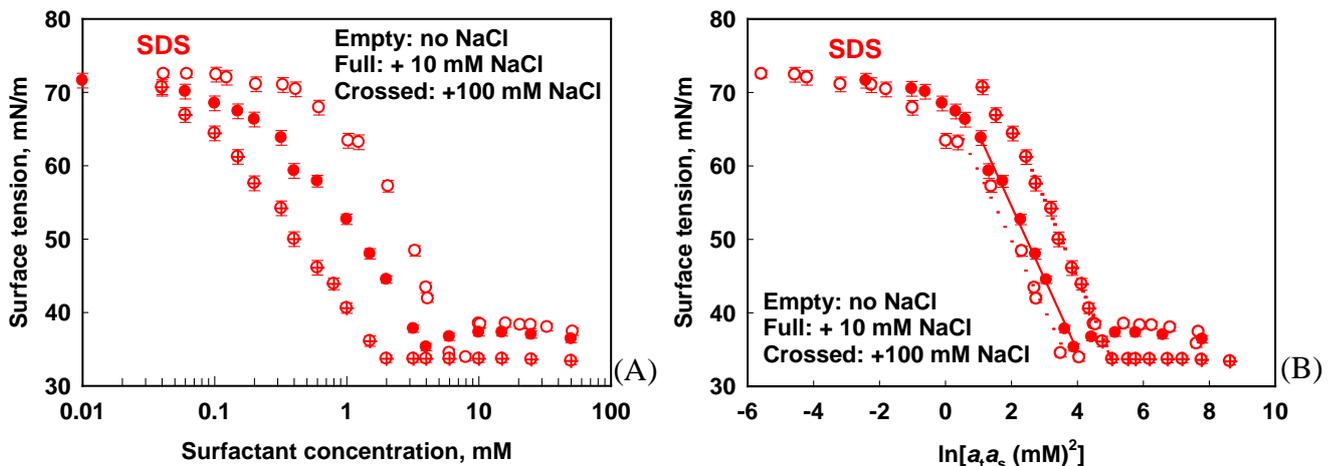

**Figure 3.** Surface tension as a function of (A) surfactant concentration and (B) $\ln[a_t a_s]$ for SDS solutions without added background electrolyte (empty symbols); with 10 mM NaCl (full symbols) and with 100 mM NaCl (crossed symbols). The curves in (B) are fits by eq. 10.

For the ionic surfactants we used the procedure proposed in Ref. [55] to determine the surfactant adsorption at CMC which consists of the following steps:

(1) We determine the activity coefficients at each surfactant and electrolyte concentrations;

(2) We assume that the excess of surfactant in the diffuse part of the electric double layer can be neglected, as shown in Ref. [55];

(3) We determine the total concentration of the counterions in the bulk which come from the surfactant and the background electrolyte;

(4) By plotting the surface tension as a function of the total counterion activity, multiplied by the surfactant activity, we determine the adsorption at CMC;

(5) To determine (approximately) the limiting adsorption we use again the Volmer model, eq. 5, as in the case of nonionic surfactants.

Below we present explicitly the equations used to realize the above procedure of data interpretation.

The mean activity coefficient is determined using the semi-empirical formula of Debye-Huckel theory which accounts for the finite size of the ions [55-57]:

$$\lg \gamma_{\pm} = -\frac{A\sqrt{I}}{1+Bd_i\sqrt{I}} + bI \tag{8}$$

Here $I$ is the total ionic strength, while the values of the constants are $A = 0.5246$ M$^{-1/2}$, $Bd_i = 1.316$ M$^{-1/2}$, $b = 0.055$ M$^{-1}$ for NaCl solutions at 20 °C.

The total ionic activity and surfactant activity are calculated by the equations:

$$a_t = \gamma_{\pm}(C_S + C_{EL}) \qquad a_S = \gamma_{\pm} C_S \tag{9}$$

Here $C_S$ is the surfactant concentration and $C_{EL}$ is the concentration of the additional inorganic electrolyte.

The experimental data for $\sigma(C_S)$ are plotted versus $\ln(a_t a_S)$. The latter dependence is fitted with the linear dependence around CMC:

$$\sigma = z_0 + z_1 \ln(a_t a_s). \tag{10}$$

The total surfactant adsorption at CMC is determined from the equation [55]:

$$d\sigma = -\Gamma_{tot} RT d\ln(a_{tot} a_S) \tag{11}$$

The values of $\Gamma_{tot}$ obtained via eq. (11) are introduced into eq. (5) to determine the respective values of the average area per molecule, $\alpha$.

The results from the above analysis are shown in Table 1 and they are in a reasonably good agreement with the experimental results reported in literature [43-44, 61-72]. Exception is the solution of DTAB ± NaCl. For the latter systems we observed a deep minimum in the surface tension isotherm around the CMC which indicates the co-adsorption of nonionic components. As a result, the total surface adsorption is higher as compared to the values determined in the literature with pure DTAB. The area per molecule in this layer, $\alpha \approx 0.23$ nm$^2$, is very close to the cross-sectional area of the hydrocarbon chain $\approx 0.21$ nm$^2$. Most probably, the nonionic component is a fatty alcohol or acid, remaining from the industrial DTAB synthesis. It is known [58-60] that the long-chain fatty alcohols and acids form dense adsorption layers with low surface tension ($\approx 22$ mN/m) as observed with this DTAB sample.

Thus we conclude that all our experimental results are in agreement with the values reported in the literature, after accounting for the presence of nonionic components in the commercial sample of DTAB.

**Table 1.** Surface properties of the studied solutions (experimental results and literature data).

| Surfactant | Experimental results | | | | Literature data | | | |
|---|---|---|---|---|---|---|---|---|
| | CMC, mM | $\sigma_{CMC}$, mN/m | $\Gamma_{CMC}$, μmol/m$^2$ | α, Å$^2$ | CMC, mM | $\sigma_{CMC}$, mN/m | $\Gamma_{CMC}$, μmol/m$^2$ | α, Å$^2$ |
| **Brij 35 + 10 mM NaCl** | 0.05 | 43.2 | 1.8 | 77 | 0.030 [46] 0.078 [47] 0.090 [48] | 42.0 [46] 43.0 [47] | 1.65 [48,47] | 88 [49,50] |
| **Brij 58 ± 10 mM NaCl** | 0.003 | 43.2 | 3.0 | 41 | 0.0028 [46] | 41.2 [51] | 2.7 [51] | 61 [49] |
| **Tween 20** | 0.012 | 38.4 | 3.6 | 35 | 0.011 [46] 0.060 [52] | 33.0 [46] 38.5 [52] | 3.05 [52] | 54.4 [52] |
| **Tween 40** | 0.022 | 42.6 | 2.3 | 59 | 0.067 [46] 0.027 [53] 0.030 [52] | 43.0 [52] | 3.0 [52] | 55.3 [52] |
| **SDS** | 8.0 | 33.9 | 3.8 | 33 | 8.2 [61-64] | 30.0 [65] | 4.0 [44] 6.0 [44] | 35 [66] 30 [43-44] |
| **SDS + 10 mM NaCl** | 4.0 | 33.9 | 4.0 | 31 | 5.0 [66-67] | 37.0 [66] | 4.6 [66] | 32 [66] 30 [44] |
| **SDS + 100 mM NaCl** | 2.0 | 33.7 | 4.2 | 30 | 1.5 [67] | 30.0 [67] | 4.3 [67] | 30 [44] |
| **DTAB** | 10 | 34.7 | 5.3 | 21 | 10 [61] | 40.0 [61] | 3.3 [61] | 37.8 [43] 36.5-39.5 [68-71] |
| **DTAB + 10 mM NaCl** | 3.2 | 25.2 | 5.8 | 21 | - | - | - | 37.8 [43] 36.5-39.5 [68-71] |
| **CTAB** | 0.82 | 37.8 | 2.9 | 46 | 0.98 [72] | 35.0 [72] | 3.0 [72] | 37.8 [43] 36.5-39.5 [68-71] |
| **CTAB + NaCl** | 0.27 | 40.0 | 2.9 | 45 | 0.15 [72] | 36.0 | 4.4 [72] | 37.8 [43] |

### 3.2. Dynamic surface tension.

To obtain information about the dynamic surface properties of the non-equilibrium adsorption layers, formed on the bubble surfaces during foaming, we measured the dynamic surface tension of the solutions studied. The concentration range between 0.1 mM and 100 mM surfactant was covered in these experiments.

The obtained experimental results are treated in the following way: (1) We calculate the universal surface age of the bubble surface using the approach from Ref. [73]. (2) The experimental data for the dynamic surface tension are fitted by eq. (13) shown below and from the best fit we determine the characteristic time for surface tension decrease, the initial surface tension, and the equilibrium surface tension; (3) Assuming that in each moment we have a unique relation between the surface tension and surfactant adsorption, as presented by eq. (5), from the measured dynamic surface tensions we determine the respective dynamic surfactant adsorption as a function of time, $\Gamma(t)$; (4) The data for $\Gamma(t)$ are fitted with a model based on the assumption for diffusion-controlled adsorption to determine the values of the initial adsorption, equilibrium adsorption and characteristic adsorption time; (5) From the parameters, determined in this procedure, we calculate the surface tension, surfactant adsorption, surface elasticity and surface coverage after 2 and 10 ms of (universal) surface age, which are used in the next section to analyze the results from the foaming tests.

The above procedure is based on the following series of equations. The universal surface age is determined by the expression proposed in Ref. [73]:

$$t_u = t_{age}/\lambda^2 \qquad (12)$$

Here $t_{age}$ is the nominal surface age, as indicated by the MBPM tensiometer, $t_u$ represents the universal surface age which does not depend on the specific tensiometer, and $\lambda^2$ is an apparatus constant which removes the effect of the bubble surface expansion during the MBPM measurements on the dependence $\sigma(t)$ which depends on various characteristics of the specific instrument. As shown in Ref. [73], $\lambda$ can be expressed via explicit integrals over the apparatus function which represents the dependence of the bubble surface area on time. $\lambda$ is independent of the surfactant type and concentration but depends on the specific MBPM apparatus. For our MBPM tensiometer this constant was determined as $\lambda^2 \approx 37$ [73].

From physicochemical viewpoint, the main difference between $t_{age}$ and $t_u$ is that $t_{age}$ corresponds to the actual lifetime of the bubbles at the tip of the capillary which releases the bubbles in the MBPM, whereas $t_u$ corresponds to an imaginary bubble with constant surface area (in contrast to the expanding area of the real bubbles in the MBPM) which would obtain the same surface tension after $t_u$. From eq (12) we see that the dependence of the surface tension on the universal surface age can be found simply via dividing the time $t_{age}$ (given by the apparatus) by 37.

The obtained experimental data were fitted by the following equation which describes very well the experimental data for not-too-small values of $t_u$ (viz. for not-too-low surface coverage) [73]:

$$\sigma = \sigma_{eq} + \frac{s_\sigma}{a_\sigma\left(1+\sqrt{t_u/a_\sigma^2}\right)} \quad (13)$$

Here $\sigma_{eq}$ is the equilibrium surface tension, $a_\sigma^2$ is the characteristic time for surface tension decrease for fixed surface area (i.e. for non-expanding bubble), and $s_\sigma$ is a parameter which accounts for the difference between the initial and the equilibrium surface tension [73].

To determine the main characteristics of the dynamic adsorption layer, formed in the process of bubble generation, we assume that the surfactant adsorption $\Gamma(t)$ can be determined from the measured dynamic surface tension $\sigma(t)$ using eq. (5). Most of the studied surfactant concentrations are around and above the CMC. Therefore, we assume that the initial surfactant adsorption is controlled by surfactant diffusion and linear relation between $\Gamma(t)$ and the subsurface surfactant concentration $C(z = 0, t)$. Under these assumptions we fit the data for $\Gamma(t_u)$ using the expression [43]:

$$\Gamma = \Gamma_{eq} + \left(\Gamma(0)-\Gamma_{eq}\right)\exp\left(\frac{t_u}{t_\Gamma}\right)\mathrm{erfc}\left(\sqrt{\frac{t_u}{t_\Gamma}}\right) \quad (14)$$

Here $\Gamma_{eq}$ is the equilibrium adsorption, $\Gamma(0)$ is the initial adsorption at $t=0$ and $t_\Gamma$ is the characteristic adsorption time. For diffusion control and surfactant concentrations below the CMC, this characteristic time is defined as [43]:

$$t_\Gamma = \frac{1}{D}\left(\frac{\partial \Gamma}{\partial C}\right)^2\bigg|_{C_{eq}} \quad (15)$$

where $D$ is the surfactant diffusion coefficient in the aqueous phase. The above equation is strictly valid only for non-ionic surfactants, whereas for ionic surfactants there is an additional electrostatic repulsion which can slow down the rate of adsorption, but for comparison we use eq. (14) for both ionic and non-ionic surfactants.

In Figure 4 we show illustrative examples for the fits of the experimental data for $\sigma(t_u)$ and $\Gamma(t_u)$ by eqs. (13) and (14), respectively. One sees that these equations describe very well the experimental data and the regression coefficients are > 0.99 for most systems.

The experimental data for the equilibrium surface tension, determined from the MBPM experiments, are in a good agreement with the results obtained by Wilhelmy plate method for most of the surfactants studied. The equilibrium adsorptions determined from the best fit to the data for $\Gamma(t_u)$ are also very close to the values of $\Gamma_{eq}$, determined from the surface tension isotherms measured by Wilhelmy plate method. Exceptions are DTAB ± 10 mM NaCl which contain significant amounts of nonionic admixtures, as explained above.

These admixtures adsorb slowly on the solution surface, especially below the CMC, and lead to lower equilibrium surface tension measured by Wilhelmy plate method, as compared to the tensions determined from the best fit to the dynamic MBPM data.

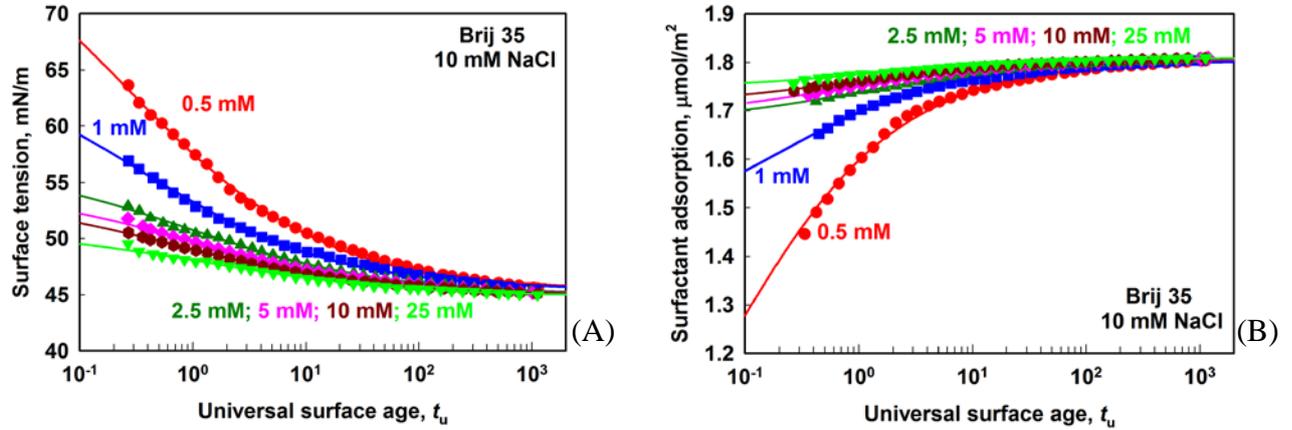

**Figure 4.** (A) Surface tension and (B) surfactant adsorption as a function of the universal surface age, $t_u$, for Brij 35 solutions at different concentrations, as shown in the graphs. All solutions contain 10 mM NaCl. The universal surface age $t_u$ is shown in ms.

The main surface characteristics of importance for the foamability of the surfactant solutions are the dynamic (instantaneous) surfactant adsorption and the related surface elasticity, surface coverage, $\Gamma/\Gamma_\infty$, and the ratio between the instantaneous adsorption and the equilibrium adsorption at CMC, $\Gamma/\Gamma_{CMC}$. These characteristics are used in Section 4 below for analysis of the results from the foam tests.

The instantaneous surface elasticity can be calculated using the following equation:

$$E_G = k_B T \Gamma_\infty \frac{\theta}{(1-\theta)^2} \tag{16}$$

where $k_B$ is Boltzmann constant, $T$ is temperature, and $\theta(t)$ is the surface coverage:

$$\theta(t) = \Gamma(t)/\Gamma_\infty = \alpha \Gamma(t) \tag{17}$$

while $\Gamma_\infty = 1/\alpha$ is the maximum adsorption in the equilibrium adsorption layer (Table 1).

The results for the various dynamic surface parameters, determined for a surface age $t_u$ = 10 ms are shown in Figure 5, as a function of surfactant concentration, $C_S$. The same results are shown in Figure S1 as a function of the normalized surfactant concentration $C_S$/CMC. The comparison of the results for the various surfactants reveals the following trends.

Dynamic surface tension, DST, for C12 nonionic surfactants (Brij 35 and Tween 20) is lower as compared to the dynamic surface tension for C16 nonionic surfactants (Brij 58 and Tween 40). For all nonionic surfactants the studied concentrations are at least 10 times above CMC, which means that the controlling factors for dynamic surface tension are the monomer concentration and the monomer release from the micelles [74]. The lowest DST for nonionic surfactants is measured for Brij 35, which has the highest CMC and, consequently, the monomer concentration is much higher for this surfactant above the CMC as compared to the other surfactants studied. In addition, the monomer release from the micelles is known to be faster for the surfactants with shorter chain at the same head group. It is worth to note also the plateau in the dependence surface tension vs. time for Brij 58 during the first 60 ÷ 400 ms (depending on the concentration) – no such plateau is observed for Brij 35. This plateau indicates that the rate of Brij 58 adsorption in the first period of surface formation is controlled predominantly by the kinetics of de-micellization, as shown in [75].

On the other hand, the dynamic surface tension for ionic surfactants with 12 C-atoms in the hydrophobic tail have much higher dynamic surface tension as compared to CTAB, which has 16C atom in the tail. The effect is very prounced in the lower concentration range (below 2 mM), see Figure 5B. The two ionic surfactants with tails of 12 C-atoms (SDS and DTAB) do not lower significantly the surface tension below the value of pure water, $\sigma_0$, in the range of low surfactant concentrations. On the other hand, the nonionic Brij 35, which also has 12 C-atoms in the hydrophobic chain, lowers $\sigma$ down to 50 mN/m within 10 ms. This slower adsorption of the ionic surfactants could be (at least partially) attributed to the pronounced electrostatic repulsion with the already adsorbed molecules of SDS and DTAB which is missing in the systems of the nonionic surfactant. The addition of 10 mM electrolyte partially suppresses the electrostatic repulsion and leads to faster decrease of the surface tension for SDS and DTAB.

CTAB has the same hydrophilic head but longer hydrophobic tail than DTAB. For these two homologues, CTAB adsorbs faster on the interface. The energy of adsorption and the molecule size (both bigger for CTAB) affect the kinetics of adsorption in opposite ways. The bigger molecules have smaller diffusion coefficient and, therefore, would adsorb slower under otherwise equivalent conditions (for diffusion-controlled and mixed regimes of adsorption). On the other hand, higher adsorption energy would lead to shorter characteristic distance and faster adsorption for barrier-controlled and mixed adsorption. The comparison between CTAB and DTAB shows that the energy of adsorption is more significant than the molecule size for these two ionic surfactants – the surfactant with longer chain adsorbs faster.

The calculated values of the surfactant adsorption, as a function of surfactant concentration (Figure 5C,D), show that the adsorption of the nonionic surfactants is the highest for Tween 20, followed by Brij 58, Tween 40 and Brij 35. Note that these adsorptions correspond to different surface coverages, $\theta = \Gamma/\Gamma_\infty$, due to the different maximum adsorptions, $\Gamma_\infty$, that can be reached by these surfactants. Therefore, the Gibbs elasticity which depends very strongly on $\theta$ (see eq. 16) is very high for Brij 35, whereas the other three surfactants have much lower elasticities.

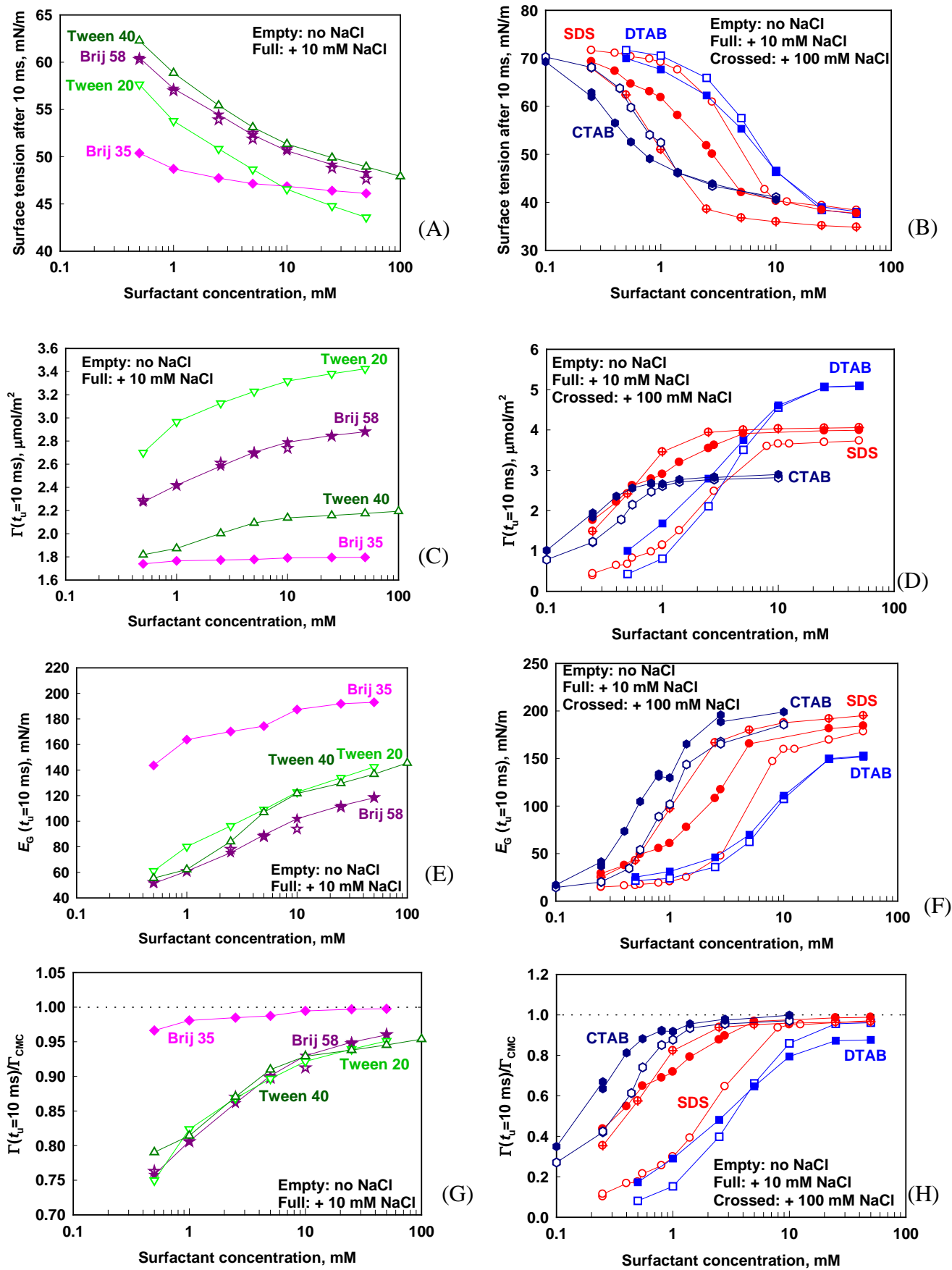

**Figure 5.** (A,B) Dynamic surface tension at 10 ms; (C,D) Surfactant adsorption at 10 ms; (E, G) Gibbs elasticity at 10 ms; (G,H) Surface coverage, $\Gamma/\Gamma_{CMC}$, vs. surfactant concentration for nonionic surfactants (A,C,E,G) and ionic surfactants (B,D,F,H) without electrolyte (empty symbols), with 10 mM NaCl (full symbols) and with 100 mM NaCl (crossed symbols).

For the ionic surfactants, the dependence of surfactant adsorption on surfactant concentration contains two distinct regions. Below the CMC, the dynamic surfactant adsorption increases almost linearly with $\ln C_S$ and remains almost constant afterwards, at a value which is very close to the value determined from the surface tension isotherm, $\Gamma_{CMC}$. The latter result means that the ionic surfactants form (almost) equilibrium adsorption layers at these higher concentrations. This can be seen also from the data presented in Figures 5G,H and in Figure S1 where the ratio $\Gamma(t_u = 10$ ms$)/ \Gamma_{CMC}$ are shown. For the ionic surfactants, $\Gamma_{CMC}$ is reached at concentrations $\approx 5 \times$CMC, except for DTAB + 10 mM NaCl for which $\Gamma$ remains of around $0.9 \times \Gamma_{CMC}$ at concentrations as high as $10 \times$CMC.

In the nonionic systems, Brij 35 reaches $\Gamma \approx \Gamma_{CMC}$ within 10 ms at concentrations above 10 mM, whereas for the other nonionic surfactants the maximum value of $\Gamma$ is up to $0.95 \times \Gamma_{CMC}$ even at surfactant concentration of 100 mM. As expected, the neutral electrolyte NaCl has no any noticeable effect on the properties of the solutions of nonionic surfactants, whereas it accelerates significantly the adsorption of the ionic surfactants.

From these results we can conclude that Brij 35 with concentrations > 1 mM is able to form equilibrium adsorption layer within 10 ms surface age, whereas the other nonionic surfactants reach $0.95 \times \Gamma_{CMC}$ at concentrations above 20 mM. All studied ionic surfactants reach surface coverage of $0.95 \times \Gamma_{CMC}$ at concentrations between 10 and 30 mM.

### 3.3. Kinetics of foam generation.

To quantify the rate of air entrapment during foaming, we measured the volume of the entrapped air, $V_A$, as a function of the number of shake cycles, see Figure 6 for illustrative results. As expected, $V_A$ increases with the increase of surfactant concentration and number of cycles (see Figure 6A). Typically, an initial fast increase of the foam volume is followed by a slower increase of $V_A$ and a plateau could be reached at large $n$. For convenience, we describe these data with the following empirical equation which contains two fit parameters with clear physical meaning:

$$V_A = V_{AMAX}(1 - \exp(-n/n_A)) \qquad (18)$$

$V_{AMAX}$ is the maximum volume of the air which would be entrapped after a very large number of cycles, $n$ is the number of the respective cycle at which $V_A$ is measured, and $n_A$ is the characteristic number of cycles at which $V_A$ reaches $\approx 63\%$ of $V_{AMAX}$. Illustrative examples of the description of our experimental data by eq 18 are shown in Figure 6A.

In Figure 6B the average liquid volume fraction in the formed foams is shown as a function of the number of shake cycles. At low surfactant concentrations, the average liquid volume fraction remains high, > 60 %, whereas it decreases down to 8 % at high surfactant concentrations and $n > 20$. There is no further possibility to decrease the average liquid volume fraction, above 8 %, because the cylinder is full with foam in the latter case.

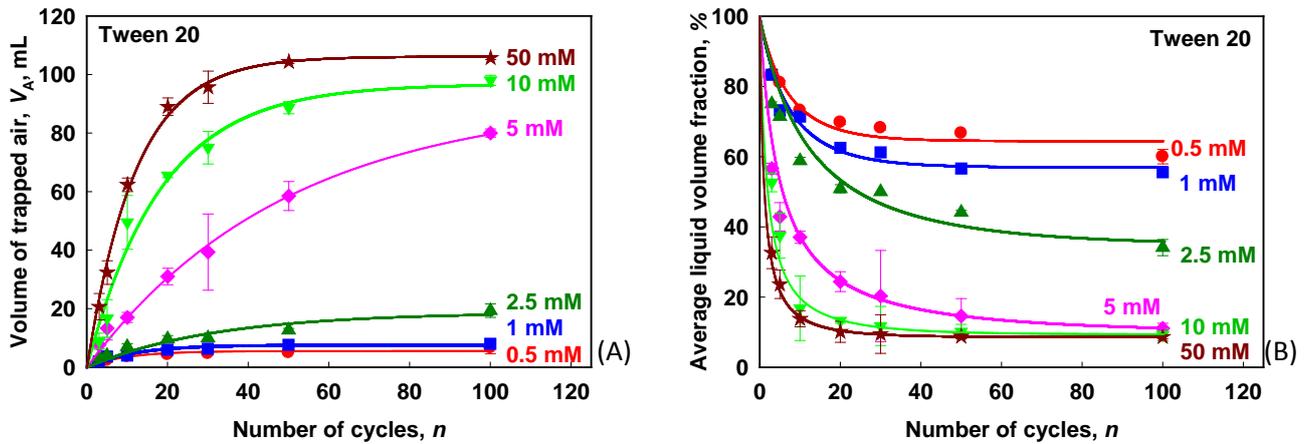

**Figure 6.** (A) Volume of the trapped air and (B) Average liquid volume fraction in the foam formed, versus the number of shake cycles for solutions of the nonionic surfactant Tween 20 at various concentrations, as shown on the graphs. The symbols show experimental data, whereas the curves are fits by eq. (18).

To characterize quantitatively the foaming process, we may use different characteristics. Equation (18) suggests the use of two characteristics which bring complementary information. The initial rate of air entrapment is characterized by the gradient $dV_A/dn|_{n\to 0} \approx V_{AMAX}/n_A$. The overall foamability of the surfactant solutions at long times is characterized by the volume of the entrapped air after 100 cycles which for most systems is $\approx V_{AMAX}$ (except for those with intermediate surfactant concentrations). These foaming parameters are compared in Figure 7 for the various surfactants, in the entire range of surfactant concentrations studied.

One sees in Figure 7 that the dependence $V_{AMAX}(C_S)$ contains 3 clearly defined regions: (1) At low surfactant concentrations $V_{AMAX}$ increases very slowly with $C_S$. In this region, the volume of entrapped air is below 20 mL. (2) Intermediate region in which $V_{AMAX}$ sharply increases with the surfactant concentration, from 20 up to 100 mL, within a 3-fold increase of concentration; (3) Plateau region in which $V_{AMAX}$ remains almost constant around 120 mL. The latter value is determined by the volume of the cylinder used in the foam test – the maximum amount of air which can be entrapped is around 120 mL (added to the 10 mL surfactant solution present in the cylinder).

To clarify how $V_{AMAX}$ depends on the CMC of the various surfactants, we plot in Figure S2 the volume $V_{AMAX}$ versus $C_S$/CMC. One sees that the transition from region 1 to region 2 occurs at a concentration of around CMC/10 for the solutions of the ionic surfactants SDS, CTAB and DTAB, while the same transition occurs at much higher values of $C_S$/CMC for the nonionic surfactants. This comparison confirms our conclusion [76] that the relative surfactant concentration, $C_S$/CMC, cannot be used as a characteristic for solution foaminess.

The initial rate of foaming $dV_A/dn|_{n \to 0}$ also exhibits 3 regions: (1) At low surfactant concentration the foaming rate is very low, ≈ 0.5 mL/cycle; (2) At intermediate concentrations we observe a rapid increase from 1 to 10 mL/cycle; (3) In the range of high surfactant concentrations, the maximal value ≈ 10 mL per cycle remains almost constant with the further increase of concentration.

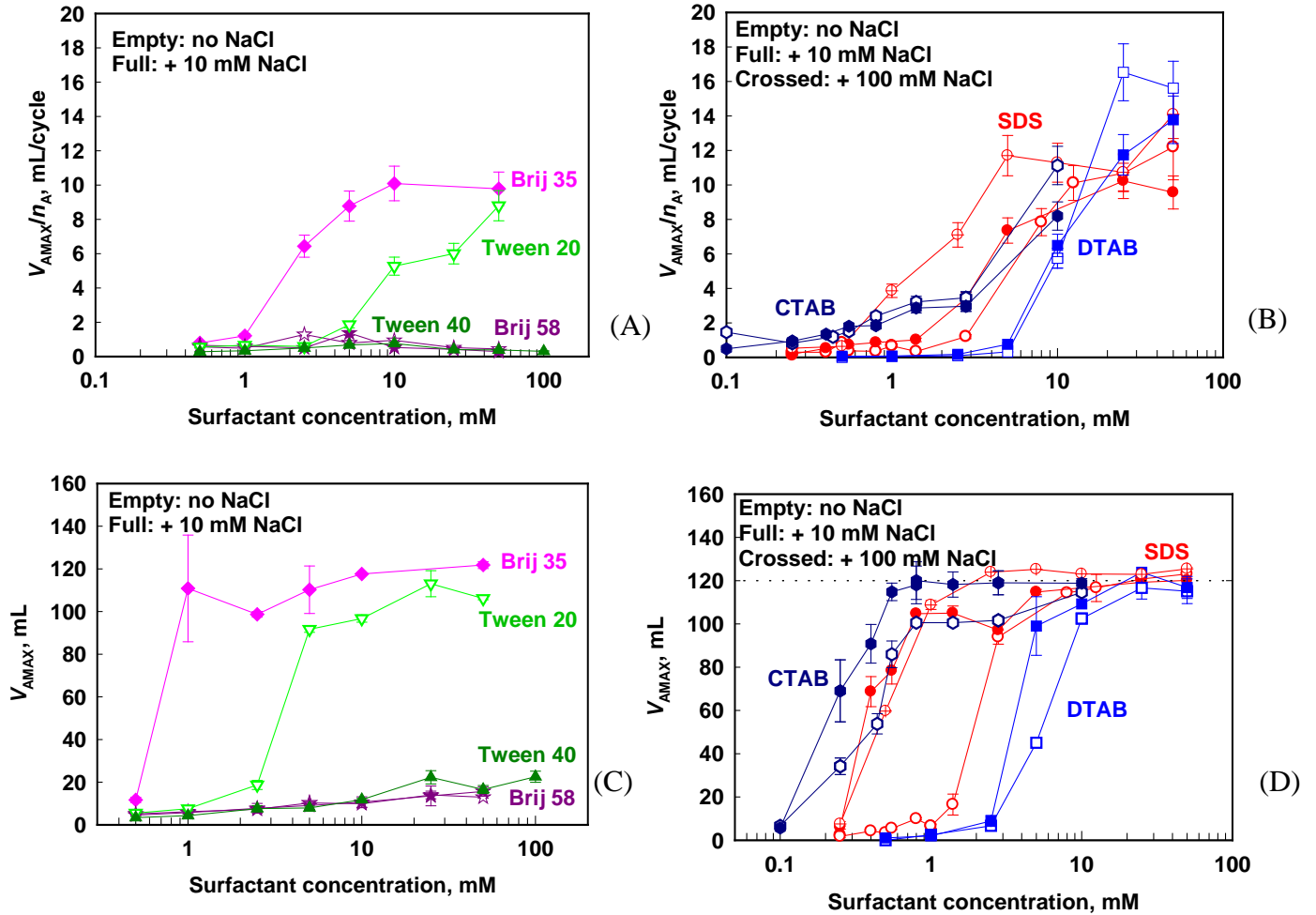

**Figure 7.** (A,B) Initial rate of air entrapment, and (C,D) maximum volume of trapped air for (A,C) nonionic and (B,D) ionic surfactants, as functions of surfactant concentration.

From these series of experiments we can conclude that nonionic surfactants with 16 C-atoms in the hydrophobic tail are not able to stabilize the bubbles in this method and, as a consequence, the amount of formed foam is very low of ≈ 20 mL, whereas the foamability of nonionic surfactants with 12 C-atoms is much better and they are able to form foam with volume of ≈ 100 mL at high surfactant concentrations. The foamability of ionic surfactant with 16 C-atoms is much better as compared to the foamability of surfactants with 12 C-atoms at low surfactants concentrations and becomes comparable at concentrations above the CMC.

## 4. Data interpretation and discussion.

In the current section we systematically check how the foaming data (initial rate and maximum foaming) correlate with those characteristics of the dynamic adsorption layers which have clear physical meaning and play a role in the processes of foam film thinning and stabilization. The major aim of this effort is to identify those key parameters which are able to explain all available experimental data on foaming, shown above. If such key parameter(s) are identified, they can provide predictive power for other systems and, furthermore, can be used as a basis for constructing a detailed theoretical model which captures all important phenomena. The latter task is rather complex and goes beyond the aims of the current study.

### 4.1. Characteristic time of the foaming process.

To construct appropriate correlation plots we need to choose a characteristic surface age which represents the specific foaming method. In the following analysis we use the dynamic quantities corresponding to universal surface ages between $t_u = 2$ ms and $t_u = 10$ ms. This range corresponds to a bubble surface age, $t_{age} = 37 t_u$, between ca. 75 and 370 ms in the MBPM. This wide range of bubble surface ages was chosen for two main reasons:

First, the optical observations of the foaming process showed that the onset of bubble coalescence in the used foaming test is observed within ≈ 60 ms after the entrapment of a new portion of air in each shake cycle. Due to the fact that there is a significant expansion of the air-water interface in the process of bubble generation in the foaming method used, we take as a lower boundary the characteristic surface age of the bubbles in the MBPM method ($t_{age} ≈ 60$ ms) and the related value of $t_u ≈ 2$ ms. On the other hand, the bubble coalescence could continue until the next shaking cycle is initiated which corresponds roughly to half of the shaking period, viz. to ≈ 600 ms. Thus, with the chosen range of values 2 ms ≤ $t_u$ ≤ 10 ms we cover the range of the surface ages of main interest for the used foaming test 75 ≤ $t_{age}$ ≤ 370 ms.

Second, the data analysis showed that all main results and conclusions remain unaffected in this entire range of surface ages – only the experimental points on the graphs shift slightly up or down. In other words, all main results and conclusions of the current study are robust with respect to the specific choice of the values of $t_u$ and $t_{age}$ if the latter fall in the range characterizing the specific foaming method.

We use the universal surface age, $t_u$, in the following analysis for two main reasons:

(1) To work with parameters which characterize the surfactant solution only and do not depend on the specific MBPM instrument used for measuring $\sigma(t)$. This would not be the case if $t_{age}$ is used, as explained in Section 3.2 and in Ref. [73];

(2) To be able in the future to check directly whether the approach and the final conclusions of the current analysis would be applicable to results from other foaming methods in which the bubble dynamics could be very different. Indeed, a continuation of the current study is under preparation in which the same surfactants and the same approach to data analysis are used to interpret results from other foaming methods. Interestingly, we found that the foaming trends observed in the other tests could be very different and these differences could be explained by considering properly for the different surface dynamics, accounted for by different values of $t_u$ for each foaming method.

### 4.2. Correlation between foam volume and dynamic surface tension.

Dynamic surface tension is sensitive to the rate of surfactant adsorption. Therefore, this parameter is often suggested as a key parameter to characterize foaming.

To check whether the observed variations in the solution foamability could be explained with differences in the dynamic surface tension of the respective surfactant solutions, we plot in Figure 8 the dependences of $V_A(n=10)$ and $V_A(n=100)$ on the DST at $t_u = 10$ ms. The complementary graphs for $dV_A/dn$ and $V_{AMAX}$ vs. DST at $t_u = 10$ ms are shown in Figure S3 of the Supplementary information. For comparison, the respective graphs for DST at $t_u = 2$ ms are shown in Figure S4 of the Supplementary information.

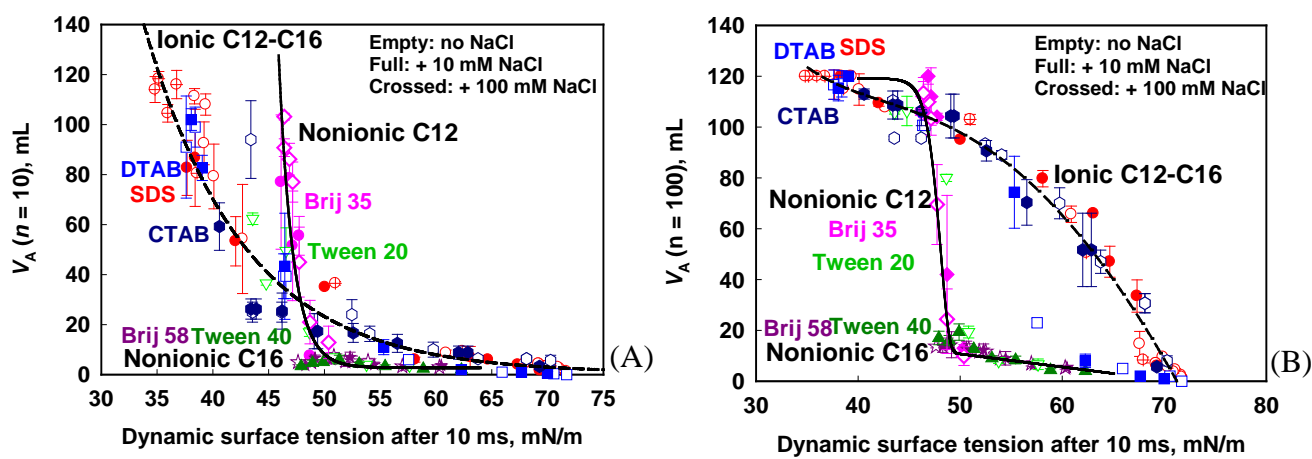

**Figure 8.** Correlation between the foaming parameters and the dynamic surface tension at $t_u = 10$ ms: (A) Volume of entrapped air after 10 shake cycles; (B) Volume of entrapped air after 100 shake cycles. Note the different shapes of these curves and the different values of DST at which the foaminess increases: for the nonionic surfactants with 12 C-atom chains DST ≈ 50 mN/m, for ionic surfactants a steep increase of the accumulated foam is seen at DST < 70 mN/m followed by a more gradual increase of the initial foaming at DST < 60 mN/m.

One sees in Figure 8 a reasonable correlation between $V_A$ and DST for the solutions prepared with the same surfactant at different concentrations. However, the results obtained with surfactants of different types do not merge around a single master curve. The comparison between the different systems shows that usually the solutions of the ionic surfactants SDS and CTAB generate larger foam, compared to the nonionic Brij 35, Brij 58, Tween 20, Tween 40 at the same value of DST. In other words, solutions containing nonionic surfactants have usually lower foamability at the same dynamic surface tension. An important exception is observed with the solutions of C12-chain nonionic surfactants Brij 35 and Tween 20 which abruptly increase their foaminess when the DST becomes < 50 mN/m – an effect which is explained below, after discussing the other correlation plots. At DST ≈ 45 mN/m these two nonionic surfactants have foaminess which is even somewhat higher than the foaminess of the solutions of ionic surfactants with the same DST.

The (generally) higher foaminess, defined as the volume of entrapped air after 100 shaking cycles, of the solutions of ionic surfactants is further reinforced by the fact that their DST could reach much lower values, down to ca. 35 mN/m, whereas the nonionic surfactants have DST ≥ 45 mN/m even at very high surfactant concentrations. Therefore, the solutions of ionic surfactants produce foam of around 120 mL when their dynamic surface tension is below ca. 40 mN/m, whereas the foams produced by nonionic surfactants are limited to 100 mL due to their higher DST. The addition of 10 and 100 mM of NaCl to the foaming solutions affects both the DST and foaminess of the solutions of ionic surfactants but the correlation points remain around the same master curve. These results confirm the existence of a significant qualitative difference between the ionic and nonionic surfactants with respect to the foaminess of their solutions, with and without external electrolyte, Figure 8B.

The ionic surfactant DTAB shows somewhat intermediate behavior between that of the ionic and nonionic surfactants. At low surfactant concentrations, corresponding to high DST, its foaminess is low and comparable to that of the nonionic surfactants. With the increase of DTAB concentration and the related decrease of DST below ca. 60 mN/m, the foaminess of the DTAB solutions becomes similar to that of the other ionic surfactants. Most probably, this peculiar behavior is affected by the noticeable fraction of nonionic admixture present in the DTAB sample, as evidenced by the minimum observed in the surface tension isotherm.

We conclude from these results that the foaminess correlates well with DST for surfactants of the same type (ionic or nonionic of the same chain length), but differs significantly when the ionic and nonionic surfactants are compared. These differences are particularly noticeable for the volume of accumulated foam after long foaming time. There is also a noticeable difference between the nonionic surfactants with C12 and C16 chains in the region of high surfactant concentrations (10 to 100 mM) and low DST (between ca. 45 and 50 mN/m).

### 4.3. Correlation between the foamability and surface coverage.

To analyze deeper the relation between the foaminess and the properties of the dynamic adsorption layers, formed on the bubble surfaces during foaming, in subsections 4.3 to 4.5 we continue with investigation of the role of surface coverage, surface elasticity and surface mobility. These surface characteristics are not independent from each other and we discuss their relation after showing the respective correlation plots.

As seen from Figure 9, for nonionic surfactants $V_A$ is relatively low and depends very weakly on $\theta_{CMC} = \Gamma/\Gamma_{CMC}$, for $\theta_{CMC} < 0.95$, while it increases very sharply at $\theta_{CMC} > 0.95$. The latter value is reached only in the solutions of the C12 nonionic surfactants, Brij 35 and Tween 20. This result shows that there is a sharp transition value for the surface coverage of the bubbles, $\theta_{CMC} \approx 0.95$, which ensures stabilization of the dynamic foam films against coalescence and allows the entrapment of air bubbles. This very high value of the surface coverage indicates that almost complete adsorption layer of nonionic surfactant molecules is needed to ensure significant foaming of the respective solutions.

In contrast, air is entrapped and the foams are stabilized at much lower surface coverage, ca. $\theta_{CMC} > 0.4$, for the solutions of ionic surfactants. Indeed, the ionic surfactants are expected to form more stable foam films at the same surface coverage, due to the strong electrostatic repulsion between the foam film surfaces [43,77-78]. The films formed by nonionic surfactants are stabilized mainly via steric interactions which are of very short range and complete adsorption layers are needed to stabilize the foam and emulsion films [43,77-78].

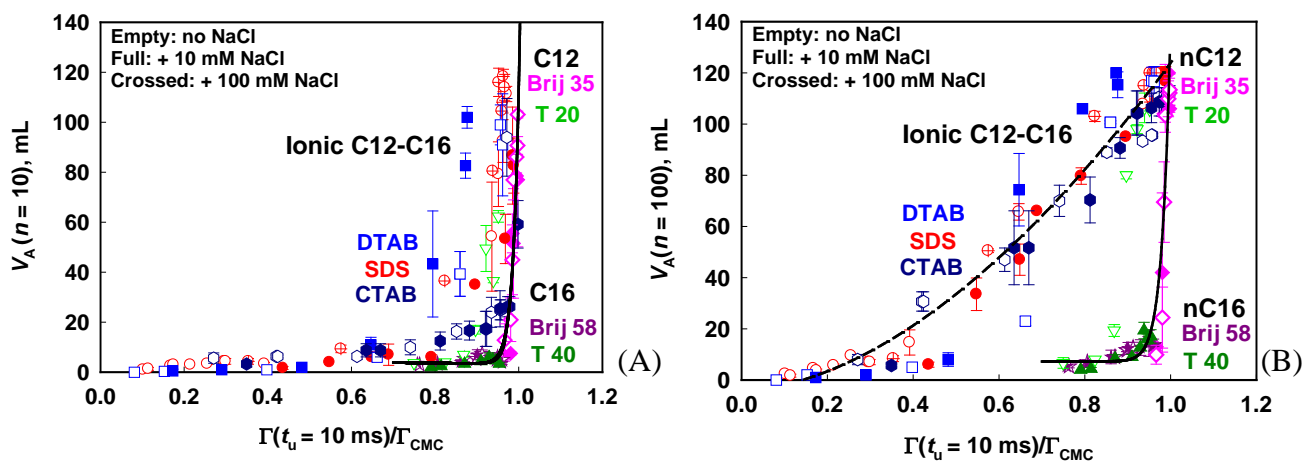

**Figure 9.** Correlation between the foaming parameters and the surface coverage, $\Gamma(t_u)/\Gamma_{CMC}$ at $t_u = 10$ ms: (A) Volume of entrapped air after 10 shake cycles; (B) Volume of entrapped air after 100 shake cycles. The curves in (B) are guides to the eye. The other measures of the initial and maximum foaming, $V_{AMAX}/n_A$ and $V_{AMAX}$, show very similar trends as shown in Figures S5 in Supplementary materials.

The data for the initial foaminess of the solutions of ionic surfactants, Figure 9A, do not merge around a single master curve – the foaminess depends not only on $\theta_{CMC}$, but also on the presence of electrolyte and the specific ionic surfactant. In contrast, the data for the accumulated foam in Figure 9B merge around two distinct master curves for the ionic and nonionic surfactants, respectively. Approximately linear increase is observed in the dependence $V_A(\theta_{CMC})$ for all ionic surfactants, with and without added NaCl, in the range 0.3 < $\theta_{CMC}$ < 1. Only DTAB has intermediate behavior - similar to nonionic surfactants at low coverage and to ionic surfactants at high coverage. As already discussed in Section 4.2, this behavior is most probably related to the high quantity of nonionic additives in this system.

### 4.4. Correlation between the foamability and dynamic surface elasticity.

The surface elasticity is related to the Marangoni effect which, in its turn, is one of the key factors controlling the hydrodynamic boundary conditions at the foam film surfaces [2,79]. Hence, the surface elasticity affects the surface mobility and the rate of foam film thinning.

For this reason, in Figure 10 we present plots of the initial and the long term foaminess of the surfactant solutions versus the Gibbs elasticity of the dynamic adsorption layers, formed at $t_u$ = 10 ms. For brevity, we call this characteristic of the dynamic adsorption layer "the dynamic Gibbs elasticity" which is not perfectly precise term but saves space and time when used.

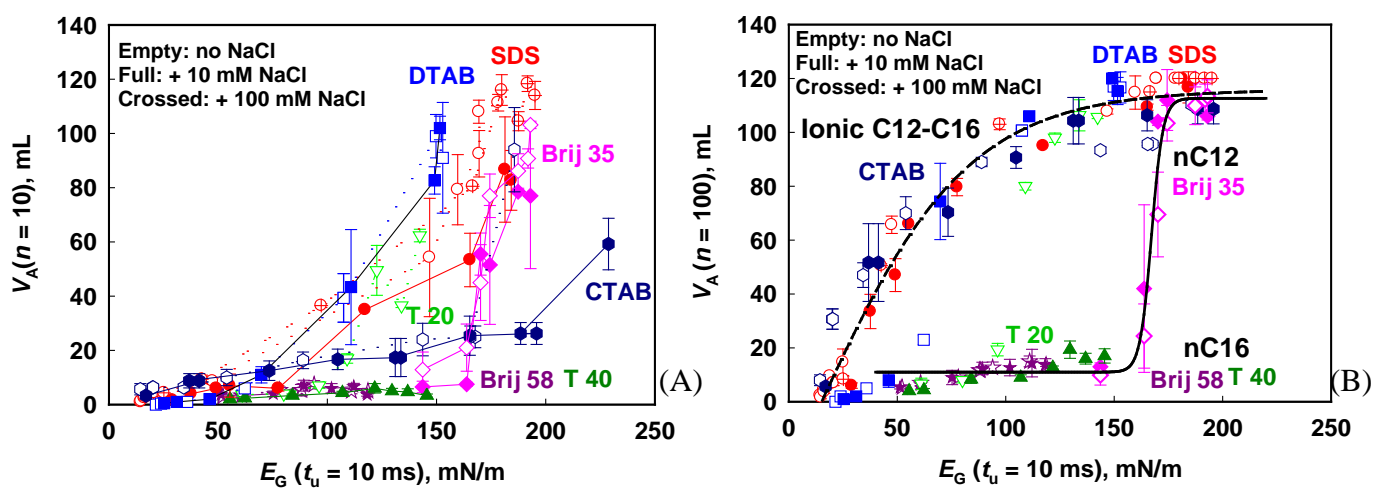

**Figure 10.** Correlation between the foaming parameters and the dynamic Gibbs elasticity at universal surface age $t_u$ = 10 ms: (A) Volume of entrapped air after 10 shake cycles; (B) Volume of entrapped air after 10 shake cycles. The other measures of initial and maximum foaming, $V_{AMAX}/n_A$ and $V_{AMAX}$, show very similar trends as seen in Figures S6 in Supplementary materials.

As seen from Figure 10B, for all nonionic surfactants we observe a very low foaminess until the dynamic Gibbs elasticity reaches values of 150 mN/m. Sharp increase of the foam volume is observed at higher Gibbs elasticity for Tween 20 and Brij 35. Somewhat surprisingly, no any correlation is seen between the initial foaminess and the dynamic Gibbs elasticity for ionic surfactants – each surfactant has very different behavior.

In contrast, all data for the accumulated foam pack very well around two master curves for the ionic and nonionic surfactants, respectively, Figure 10B. The addition of NaCl up to 100 mM does not change the type of behavior of the ionic surfactants – all data are grouped very well around the respective master curve. The latter result shows that this high electrolyte concentration does not suppress completely the electrostatic repulsion between the film surfaces, despite its strong quantitative effect on the range and magnitude of the electrostatic interactions. Correlations like that in Figure 10B can be used as a firm basis for development of new and rigorous theoretical models of foaming which should include both the properties of the dynamic adsorption layers and the electrostatic repulsion between the foam film surfaces during foaming.

### 4.5. Correlation between the foamability and surface mobility.

The relation between the surface elasticity (Marangoni effect) and surface mobility is well established in literature. However, the hydrodynamic flow in foam and emulsion films is strongly affected by the interactions between the film surfaces and especially by the complex processes of mass transfer of surfactant, including adsorption, bulk and surface diffusion, and bulk and surface convection. Various elaborated theoretical approaches were proposed to describe these complex processes [80-82] but all of them require extensive numerical simulations of the interplay between surfactant mass transfer and adsorption, on one side, and the hydrodynamics of film thinning, on the other side. Such a numerical effort is not justified in the current context, because the dynamics of foam film thinning is much more complex in the actual foaming process, when compared to the idealized assumptions used to develop these theoretical models.

Therefore, following our approach, in the current subsection we compare the foaminess of the surfactant solutions with surface characteristics which are shown theoretically to account for the effect of surfactants on the rate of foam film thinning. Due to the nature of the surfactants used in our study, we focus our search on parameters characterizing the low-molecular mass surfactants: bulk and surface diffusion, Marangoni effect, etc. We do not consider the effect of the real surface viscosity, as it is expected to play a role for surface active species with more complex molecular structure (e.g., saponins or polymers incl. proteins, hydrophobized polysaccharides and other natural or modified polymers) which were found to form viscoelastic adsorption layers [83-84].

Parameters characterizing the effect of low molecular mass surfactants on the rate of foam film thinning were defined in the theoretical approach developed by Ivanov and collaborators [85]. These authors considered the rate of thinning of foam films with an explicit account for the effect of soluble surfactants on the surface mobility and the rate of foam film drainage. Assuming the presence of adsorption layers which are not far away from equilibrium (which is one of the serious limitations of this approach) these authors showed that the rate of foam film thinning can be approximated by the following expression:

$$\frac{V_{DR}}{V_{RE}} = 1 + b + \frac{h_s}{h} \tag{19}$$

Here $h(t)$ is the instantaneous film thickness, $b$ and $h_S$ are characteristics of the surfactant solution which account for the surface mobility, $V_{DR}$ is the rate of film drainage in the presence of surfactants in the aqueous phase, and $V_{RE}$ is the Reynolds velocity of thinning of planar film with tangentially immobile surfaces [86]:

$$V_{Re} = \frac{2F h^3}{3\pi\eta_c R_F^4} = \frac{2P_C h^3}{3\eta_c R_F^2} \tag{20}$$

$F$ in Eq. (20) is the external force, pushing the bubble against a large interface, $\eta_C$ is the viscosity of the surfactant solution, and $R_F$ is the radius of the foam film, $R_F \approx (FR_b/\pi\sigma)^{1/2}$, which is determined from the force balance. As seen from eq. (20), the driving force for film thinning, $F = \pi R_F^2 P_C$, could be expressed through the capillary pressure of the bubble, $P_C \approx 2\sigma/R_b$.

Both the experiments and theoretical modelling have shown that the foam films rupture and the bubbles coalesce (unless strong repulsive forces are able to stabilize the film) after reaching a certain critical thickness [87]:

$$h_{CR} = 0.21 \left( \frac{A_H^2 R_F^2}{\sigma P_C} \right)^{1/7} \tag{21}$$

where $A_H \approx 4.1 \times 10^{-20}$ J is the Hamaker constant for aqueous films in air. The critical film thickness is typically of the order of 30 nm for a millimeter sized foam films [8-9]. Equation (21) is derived under the assumption that the van der Waals forces prevail over all other forces. This is a very reasonable assumption for bubbles which coalesce with each other, like those in the foams studied, because the coalescence confirms that the attractive forces prevail.

Assuming that the bubble coalesces with the large interface under the buoyancy force, $F \approx \Delta\rho g V_B$, combining equations (20) and (21), one obtains the following expression for the lifetime of a film with tangentially immobile surfaces [88-89]:

$$t_{Re} = 4.1 (\Delta\rho g)^{5/7} \eta_C R_b^{25/7} A_H^{-4/7} \sigma^{-8/7} \tag{22}$$

where $\Delta\rho$ is the mass density difference between the surfactant solution and air, $g$ is the gravity acceleration, $V_B = (4/3)\pi R_b^3$ is the bubble volume, and $R_b$ is the bubble radius.

If the surfaces of the foam bubbles were immobile (blocked by the adsorbed surfactant molecules) we would have $b \ll 1$ and $h_s/h \ll 1$ in eq. (19), which corresponds to rate of film thinning described by Reynolds equation, $V_{DR} \approx V_{Re}$, and film lifetime described by eq. (22). Note that for the typical foaming solutions, like those used in the current study, $\Delta\rho \approx 10^3$ kg/m$^3$, $\eta_C \approx 1$ mPa.s and $A_H \approx 4.1 \times 10^{-20}$ J are fixed. Therefore, the specificity of the surfactants is reflected in eq. (22) only in the bubble size, $R_B$ (which could be different) and the dynamic surface tension, $\sigma$. If only these two parameters were important, one could expect that the dynamic surface tension would serve as the only governing parameter, because $R_B$ is also expected to depend primarily on the values of $\sigma$ under such conditions. However, as seen in Figure 8, no general correlation is observed between the foaming results and the dynamic surface tension. The reason is that eq. (22) neglects both the surface mobility of the foam films (which may vary significantly for the different solutions) and the surface forces (e.g. the electrostatic repulsion for ionic surfactants).

To make a step further and to include in our consideration the effect of surfactants on the surface mobility and on the rate of foam film thinning, we analyse below the role of the parameters, accounting in eq. (19) for the mobility of the foam film surfaces:

$$b = \frac{3\eta D_{BC}}{h_a E_G} \qquad\qquad h_a = \frac{\partial \Gamma}{\partial C} \qquad\qquad (23)$$

$$h_S = \frac{6\eta D_{SC}}{E_G} \qquad\qquad (24)$$

Here $D_{BC}$ and $D_{SC}$ are the bulk and surface diffusion coefficients of the surfactant molecules, $E_G$ is the Gibbs elasticity of the instantaneous (dynamic) adsorption layer, $h_a$ accounts for the surface activity of the surfactant. As shown by Radoev et al. [85], higher values of $b$ and $h_S$ correspond to faster diffusion from the film interior (characterized by $b$) or along the film surface (characterized by $h_S$). Both the bulk and surface diffusion act to restore the homogeneous distribution of surfactant molecules on the film surfaces, thus suppressing the Marangoni effect, increasing the surface mobility and accelerating the foam film drainage.

In a following study, Stoyanov and Denkov [90] revealed that the diffusion coefficients entering eqs. (23) and (24) should be the collective diffusion coefficients which include contributions from the interactions with the other surfactant molecules present in the bulk and in the adsorption layer, respectively. These interactions are particularly important for the surface diffusion of the molecules, characterized by $D_{SC}$, because the molecule density in the adsorption layer is usually high and the intermolecular interactions are very significant. In contrast, the interactions in the bulk are usually negligible due to the low surfactant concentration. Accounting for the role of interactions, the following expressions for the bulk and surface diffusion coefficients were proposed [90]:

$$D_{BC} = D_{B0}\frac{K_B(\phi_1)}{1-\phi_1} \approx D_{B0} \qquad D_{SC} = D_{S0}K_S(\Gamma_1)\frac{E_G}{kT\Gamma} \qquad (25)$$

Here $D_{B0}$ and $D_{S0}$ are the bulk and surface diffusion coefficients, respectively, of single surfactant molecules which do not interact with other surfactant molecules. Therefore, $D_{B0}$ and $D_{S0}$ do not depend on surfactant concentration and adsorption. $K_B(\phi_1)$ is the dimensionless mobility function of the surfactant molecules which accounts for the changes in the hydrodynamic friction between the fluid and surfactant molecules within the bulk and $\phi_1$ is the surfactant volume fraction in the bulk. In the systems considered here $\phi_1 << 1$, $K_B(\phi_1) \approx 1$ and $D_{BC} \approx D_{B0}$. Therefore, to use this approach we need approximate expressions for $D_{B0}$, $D_{S0}$ and $K_S(\Gamma)$. Such approximate expressions with the respective argumentation are presented in Section S2 in Supplementary material and lead to the following final equations for the parameters controlling the surface mobility of the foam films:

$$b = \frac{C}{2\pi r \frac{\text{Arccosh } p}{\sqrt{p^2-1}} \Gamma^2} \qquad h_S = \frac{(1-\theta)}{2r\Gamma} \qquad (26)$$

where $r$ is the radius of the hydrophilic headgroup, $p = r/l$ is the ratio between the length of the hydrophobic tail and the headgroup radius (determined as explained in Section S2 in Supplementary material), while $C$ and $\Gamma$ are expressed in molecules/m$^3$ and molecules/m$^2$, respectively.

One sees from eq. (26) that both diffusion-related factors (in the bulk and in the surface layer) which affect the film drainage rate, decrease strongly with the increase of surfactant adsorption, which means that the higher adsorption would lead to slower film thinning and longer drainage time.

When we tried to apply eq. (19) to account for the surface mobility of the foam films, we found that for all realistic parameters of the used surfactants and all realistic film thicknesses, $h \geq h_{CR}$, both mobility parameters $b << 1$ and $h_S/h << 1$. In other words, eqs. (19) and (26) predict that the foam film surfaces should always behave as tangentially immobile. The latter assumption could not be true and the most probable reason for this discrepancy of the theoretical model with the reality is that the model is heavily based on the assumption that the surfactant adsorption layers on the foam film surfaces are close to equilibrium. As evidenced very clearly by the dynamic surface tension data, for most surfactant solutions studied, especially those for which the bubble coalescence is important, the dynamic adsorption determined at $t_u$ is very far from the equilibrium one.

Therefore, instead of trying to apply directly eqs (19) and (26) for data interpretation, we use the fact that the two factors accounting for the surface mobility of the foam films provide convenient combinations of physicochemical parameters which probe these effects. In other words, we checked whether we can correlate the foaming results with the values of $b$ and $h_S$, as defined in eq. (26). To work with dimensionless quantities, in the correlation plots we have normalized $h_S$ with the critical thickness for film rupture, $h_{CR}$, because this is the natural film thickness when considering the drainage time and rupture of foam films – see e.g. Refs. [7,9,43,86,91-93]. Furthermore, we found that for all solutions studied, $b << h_S/h_{CR}$, which means that the surface mobility and the film thinning in these systems is governed predominantly by the diffusion of the surfactant molecules along the surfaces of the foam films (the effect of bulk diffusion is of secondary importance). Therefore, in the analysis below we consider only the effect of the surface diffusion through the values of $h_S/h_{CR}$, which express the main effect of the various surfactants on the rate of film thinning.

The respective correlation plots are shown in Figure 11. As with the other characteristics accounting for the surface mobility (dynamic Gibbs elasticity and surface coverage), no master line is seen for the ionic surfactants when the amount of generated foam after 10 cycles (a measure of the initial foamability of the solutions) is plotted versus $h_S/h_{CR}$. The data for ionic surfactants are scattered and, for a given ionic surfactant, the curves depend on the electrolyte concentration. For the nonionic surfactants, as in the previous subsections, we observe low foaming until the surface mobility drops to a certain critical value, at which the data for Tween 20 and Brij 35 rapidly jump up to high foam volume.

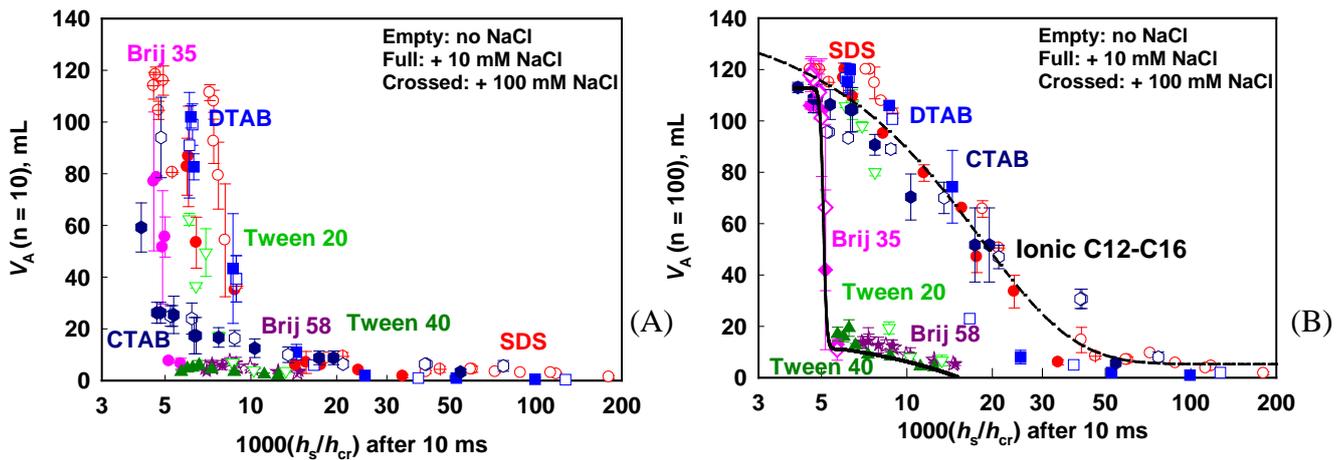

**Figure 11.** Correlation between the foaming parameters and the surface mobility factor, $h_S/h_{CR}$, at $t_u = 10$ ms, which accounts for the mobility of the film surfaces upon film thinning: (A) Volume of entrapped air after 10 shake cycles; (B) Volume of entrapped air after 100 shake cycles. The other measures of initial and maximum foaming, $V_{AMAX}/n_A$ and $V_{AMAX}$, show very similar trends as seen in Figures S7 in Supplementary materials.

In contrast, when we plot the amount of foam formed after 100 cycles against $h_S/h_{CR}$ we observed two master curves – one for the nonionic surfactants and another one for the ionic surfactant. This result confirms that the electrostatic repulsion additionally suppresses the bubble coalescence for ionic surfactants, thus facilitating the accumulation of foam at longer foaming times.

### 4.6. Comparison of the main trends and main governing factors.

In this subsection we compare the various correlation plots and, on this basis, draw conclusions about the role of the various factors studied in the foaming phenomenon.

The comparison of the correlation plots in Figures 8-11 shows that in all cases we should clearly distinguish between the initial foaming, expressed as the foam volume after 10 cycles or as the initial rate of foaming, and the foaming at long times, expressed as the foam volume after 100 cycles or as $V_{AMAX}$. Also, in all cases we observe clear quantitative difference between the nonionic and ionic (± NaCl) surfactants – the data for these two types of surfactants group around two very distinct master curves.

Excellent differentiation of the foaminess at long time, for all systems studied, is observed with the dynamic Gibbs elasticity, $E_G$ (Figure 9B), surface coverage, $\theta_{CMC}$ (Figure 10B), and film surface mobility, $h_S/h_{CR}$ (Figure 11B). All these parameters characterize the surface mobility and the related rate of film drainage. Thus we can conclude firmly that the most important key factor in the formation of voluminous foam is the surface mobility – it should be reduced significantly to ensure sufficiently long time for the surfactant to adsorb on the foam film surfaces and to stabilize the film by repulsive forces when its thickness approaches the critical film thickness, $h_{CR}$. The main difference between the ionic and nonionic surfactants is that the ionic surfactants are able to stabilize the foam films at much lower surface coverage (viz. at higher surface mobility) due to the strong electrostatic repulsion between the film surfaces, covered with charged surfactant molecules. In contrast, very high surface coverage, $\theta_{CMC} > 0.95$, related to high dynamic Gibbs elasticity, $E_G > 150$ mN/m, is needed to ensure low surface mobility and steric stabilization of the foam films. These important effects of the surface coverage and surface mobility on the foaming for ionic and nonionic surfactants are schematically illustrated and compared in Figure 12.

Note that the addition of NaCl (up to 100 mM) affects strongly the characteristics of the adsorption layers of ionic surfactants – the adsorption is faster, the DST is lower, and the foam is bigger at higher electrolyte concentration. Nevertheless, the clear differentiation between the ionic and nonionic surfactants is not affected by the addition of NaCl – the data always fall around the respective master curves, independently on the electrolyte concentration in the studied range of ionic strengths. This effect of NaCl confirms unambiguously that there is a qualitative difference between the foaming properties of the ionic and nonionic surfactants. In other words, the electrostatic repulsion in the presence of ionic surfactants ensures additional stability of the foam films which is missing at low concentrations of the nonionic surfactants.

We note that long-range electrostatic repulsion was reported for foam films of some nonionic surfactants [94-96]. This electrostatic repulsion is created by the adsorption of hydroxyl groups on the bare air-water interface [94]. However, this repulsion is unable to stabilize the bubbles during foaming in the absence of surfactants – otherwise, we would be able to generate foams without any surfactant. Increasing the adsorption of nonionic surfactants on the air-water interface was found to decrease the electrical surface potential and, thus, to suppress even further these electrostatic effect for nonionic surfactants [94]. Therefore, this effect has negligible contribution to the foaminess of the the surfactant solution.

The various correlation plots for the initial foaming show that grouping of the experimental data is observed only when the initial foaming is plotted against the dynamic surface tension (DST). In this plot, almost all experimental data group around two master curves with very different shapes – one for the ionic and another one for the nonionic surfactants. In all other correlation plots the data for the ionic surfactants are scattered and differ for the specific surfactants and NaCl concentrations. This comparison shows that the initial stage of foaming is strongly affected by the ability of the air-water interface to rapidly stretch upon agitation, as the lower DST corresponds to lower energy demand for surface stretching at the same mechanical perturbation. The different shapes of the curves for nonionic and ionic surfactants reflect the additional important effect of the stabilization of the newly formed dynamic foam films by the surface forces – electrostatic for the ionic surfactants and steric for the nonionic ones.

The initial foaming for the nonionic surfactants correlates well with all other characteristics studied: $\theta_{CMC}$, $E_G$ and $h_S/h_{CR}$. In all cases, a well-defined threshold value, ensuring rapid increase in the initial rate of foaming, is observed (e.g. $\theta_{CMC} > 0.95$ and $E_G > 150$ mN/m). The reason for these relations is that both the surface mobility rapidly decreases and the steric repulsion sharply increases above these threshold values, with the formation of dense adsorption layer of non-charged molecules.

Note that the DST is not a good discriminator for the foaming after long time (viz. for foam accumulation) neither for ionic nor for non-ionic surfactants. Indeed, the data in Figure 8B show not only a difference between the ionic and nonionic surfactants, but also a significant difference between non-ionic surfactants with 12 and with 16 carbon atoms chain-lengths. The data for DTAB also show a peculiar behaviour and deviations from both master curves in this plot.

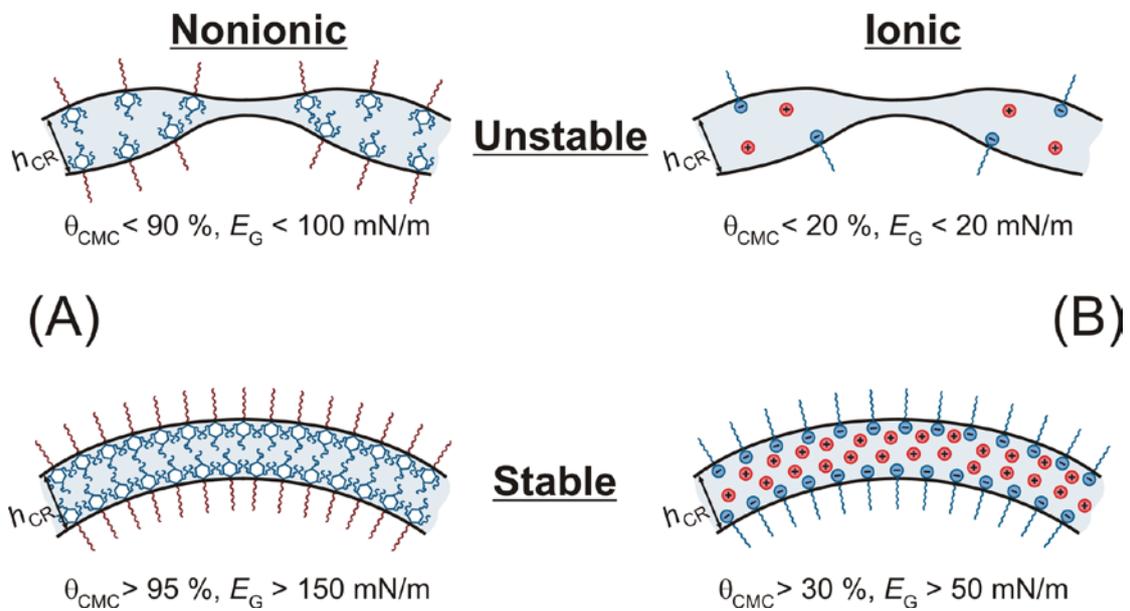

**Figure 12.** Schematic presentation of the two main phenomena, surface mobility and foam film stabilization, in dynamic films during foaming: (A) For nonionic surfactants, very high surface coverage is needed to suppress the surface mobility, reduce the rate of film thinning (thus ensuring longer time for surfactant adsorption) and stabilize the film by steric forces. (B) For ionic surfactants, the faster adsorption and the strong electrostatic repulsion ensure film stabilization at much lower surface coverage. Note that the foam films could be very inhomogeneous in thickness and surfactant adsorption during foaming. Therefore, they could break locally, at some thinner regions with lower surfactant adsorption, even if the average film thickness and the average surfactant adsorption are relatively large.

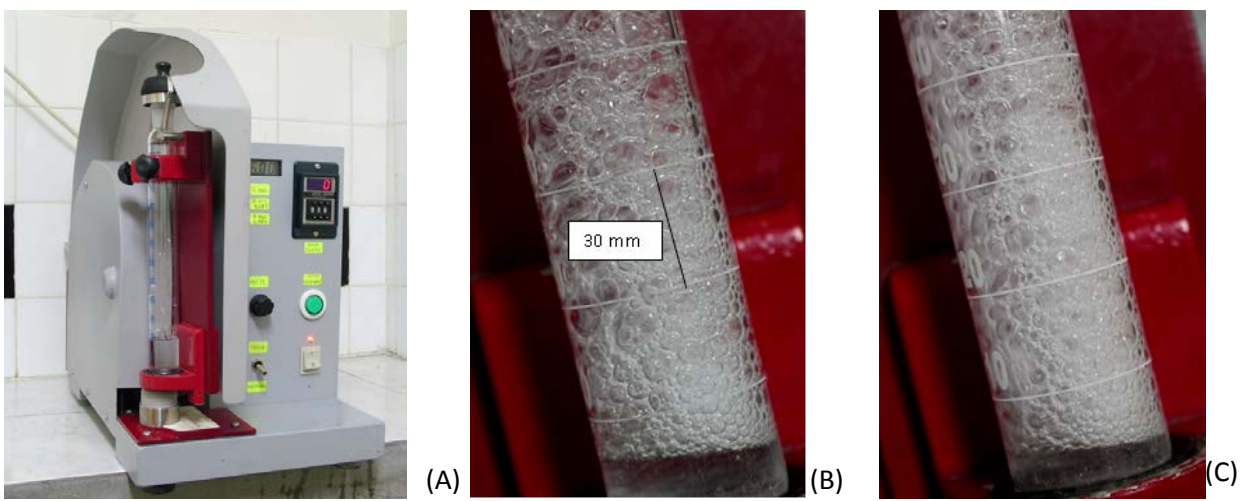

**Figure 13.** (A) Picture of the apparatus used and (B,C) Foams generated after 10 shaking cycles from (B) 10 mM Brij 35 and (C) 10 mM SDS

## 4.7. Comparison with emulsification.

In a series of previous studies [97-99] we have clarified the role of surfactant type and concentration for the efficiency of emulsification, expressed through the size of the formed emulsion drops in turbulent flow. Although very significant differences in the dynamics of foaming and emulsification could be easily identified, it is worthy to compare the effects of the surfactants on these two processes and to define clearly the similarities and differences.

In both processes, foaming and emulsification, qualitative difference was observed between the ionic and nonionic surfactants. In both processes stabilization of the bubbles and drops by nonionic surfactants was achieved only when the surface coverage approached very high values, ca. $\theta > 90$ %. These results are easily explained by considering the need of very high surface coverage to ensure steric stabilization by the non-ionic surfactants. In contrast, in both foaming and emulsification the ionic surfactants ensure bubble/drop stabilization at much lower surface coverage, due to the important contribution of the electrostatic repulsion between the film surfaces which prevents the bubble/drop coalescence.

Thus we conclude that the main phenomena and the general effect of surfactants are very similar in both types of processes. Even more important is the conclusion that the results of foaming and emulsification should be analysed by considering the dynamic adsorption layers which protect the drops and bubbles against coalescence. Simpler characteristics, such as total surfactant concentration or the scaled surfactant concentration, $C_S$/CMC, which do not account for the dynamic surface properties of the surfactants, cannot be used to explain the results of foaming or emulsification.

However, when we consider the specific surface characteristics which were successfully applied to explain the experimental data, we see some subtle differences between foam and emulsion systems. For example, emulsification in the presence of ionic surfactants was observed at extremely low surface coverage (below 1 %) [97], whereas we observed that at least 20 % surface coverage by ionic surfactants is needed to observe noticeable volume of foam in the foam test used in the current study. All these differences reflect the very different size of the drops and bubbles, as well as the very different hydrodynamic conditions in these experiments. As a result of these differences, the size of the foam and emulsion films and the hydrodynamic forces pushing the drops and bubbles differ by orders of magnitude and these dissimilarities explain the quantitative differences observed in the description of the effect of surfactants in foaming and emulsification.

Thus we conclude that the same key phenomena – the surface mobility and the surface forces in the dynamic foam and emulsion films govern the coalescence processes during foaming and emulsification, as illustrated in Figure 12. However, any quantitative analysis of these processes requires a proper account for the very different sizes of the main entities (bubbles and drops) and the different hydrodynamic conditions.

## 5. Main results and conclusions

Systematic series of experiments with seven anionic, cationic and nonionic surfactants of various molecular structures are performed. The foamability, and the equilibrium and dynamic surface tensions of the surfactant solutions are measured in wide range of surfactant and electrolyte concentrations. From the dynamic surface tension we determined the dependence of the surfactant adsorption, surface coverage, and instantaneous surface elasticity on the surface age of the bubbles, viz. along the formation of the dynamic adsorption layer during foaming.

The foaming data revealed that one should distinguish between two characteristics of the foam formation process – the initial rate of foaming and the foaming at long times (foam accumulation). These two characteristics exhibit different trends when related to the physicochemical properties of the dynamic adsorption layers and should be analysed separately. Qualitative difference was observed also between the nonionic and ionic surfactants.

All results from the foaming tests could be explained conceptually by considering two key properties of the dynamic foam films, formed during the foaming process: (1) the surface forces which could stabilize the foam films and (2) the surface mobility which affects strongly the rate of film thinning and, thus, controls the time available for surfactant adsorption before the film thins down to its critical thickness for rupture. The ionic and nonionic surfactants affect in different ways these film properties, as illustrated in Figure 12, which explains their different behaviour upon foaming.

The films formed from solutions of nonionic surfactants are stabilized via steric repulsion which becomes sufficiently high to prevent bubble coalescence only at relatively high surface coverage on the surfaces of the dynamic foam films, $\theta_{CMC} > 95\,\%$. This latter result is mechanistically similar to the observed stabilization of the emulsion drops in presence of nonionic surfactants at similarly high values of the surface coverage [94,96]. The transition is very sharp – all solutions with lower surface coverage produce small amount of foam whereas the solutions with higher surface coverage produce voluminous foam. The reason is that all key characteristics of the dynamic adsorption layers which govern the surface mobility and the steric repulsion in the dynamic foam films increase very sharply when the surface coverage approaches that of the dense adsorption layer. The related threshold value of the Gibbs elasticity of the dynamic adsorption layers is $\approx 150$ mN/m.

The data for the ionic and non-ionic surfactants merge around different master curves when plotted as a function of the various surface characteristics. This difference between the ionic and nonionic surfactants is explained with the important contribution of the electrostatic repulsion between the foam film surfaces for the ionic surfactant which additionally stabilizes the dynamic foam films during foaming. Therefore, much lower surface coverage ($\theta_{CMC} > 30\,\%$) and Gibbs elasticity ($E_G > 20$ mN/m) are sufficient to observe a noticeable foam volume for ionic surfactants. A gradual increase of the volume of trapped air, $V_A$, is observed with the increase of $\theta_{CMC}$ and $E_G$. Interestingly, $V_A$, is approximately a linear function of $\theta_{CMC}$ in a very wide range of surface coverages. All data for the ionic surfactants, without and with added up to 100 mM NaCl, group around the same master curves. The latter result supports the conclusion that the ionic surfactants have qualitatively different properties when compared to the nonionic ones.

No simple correlation is observed between the foaminess of the surfactant solution and the surfactant concentration, either total or scaled by the CMC. The dynamic surface properties, explained above, are much more important than the bulk concentration (*per se*) of the surfactant. On the other hand, the surface properties are intimately related to the bulk concentration, demicellizaton rate and other properties of the bulk surfactant solutions. In the current approach, this relation is accounted for explicitly by using the data for the dynamic surface tension at the appropriate surface age in the data analysis – this dynamic surface tension reflects the rate of surfactant adsorption and all other related properties of the bulk solution.

All observed trends with ionic and nonionic surfactants have clear physicochemical explanations and can be used as a solid basis for the development of future detailed models of foaming in systems, in which the bubble coalescence has significant contribution.

**Funding:** This work was supported by Unilever R&D Vlaardingen, the Netherlands, and by Operational Program "Science and Education for Smart Growth" 2014-2020, co-financed by European Union through the European Structural and Investment Funds, Grant BG05M2OP001-1.002-0012 "Sustainable utilization of bio-resources and waste of medicinal and aromatic plants for innovative bioactive products".

**Figure captions**

**Figure 1.** Schematic presentation of the main physicochemical processes which define the foam volume upon foaming. The foam volume is determined by the interplay between the processes of air entrapment and bubble coalescence with the large air-water interface. On its turn, the coalescence depends on the competition between the rates of surfactant adsorption and foam film thinning to the critical thickness of film rupture.

**Figure 2.** Surface tension isotherms for the studied nonionic surfactants: (A) Brij 35 and Tween 20; (B) Tween 40 and Brij 58. In all graphs in the paper, the empty symbols represent data obtained without any additional electrolyte, the full symbols present data obtained in the presence of 10 NaCl, and crossed symbols present data obtained at 100 mM NaCl.

**Figure 3.** Surface tension as a function of (A) surfactant concentration and (B) $\ln[a_t a_s]$ for SDS solutions without added background electrolyte (empty symbols); with 10 mM NaCl (full symbols) and with 100 mM NaCl (crossed symbols). The curves in (B) are fits by eq. 10.

**Figure 4.** (A) Surface tension and (B) surfactant adsorption as a function of the universal surface age, $t_u$, for Brij 35 solutions at different concentrations, as shown in the graphs. All solutions contain 10 mM NaCl. The universal surface age $t_u$ is shown in ms.

**Figure 5.** (A,B) Dynamic surface tension at 10 ms; (C,D) Surfactant adsorption at 10 ms; (E, G) Gibbs elasticity at 10 ms; (G,H) Surface coverage, $\Gamma/\Gamma_{CMC}$, vs. surfactant concentration for nonionic surfactants (A,C,E,G) and ionic surfactants (B,D,F,H) without electrolyte (empty symbols), with 10 mM NaCl (full symbols) and with 100 mM NaCl (crossed symbols).

**Figure 6.** (A) Volume of the trapped air and (B) Average liquid volume fraction in the foam formed, versus the number of shake cycles for solutions of the nonionic surfactant Tween 20 at various concentrations, as shown on the graphs. The symbols show experimental data, whereas the curves are fits by eq. (18).

**Figure 7.** (A,B) Initial rate of air entrapment, and (C,D) maximum volume of trapped air for (A,C) nonionic and (B,D) ionic surfactants, as functions of surfactant concentration.

**Figure 8.** Correlation between the foaming parameters and the dynamic surface tension at $t_u = 10$ ms: (A) Volume of entrapped air after 10 shake cycles; (B) Volume of entrapped air after 100 shake cycles. Note the different shapes of these curves and the different values of DST at which the foaminess increases: for the nonionic surfactants with 12 C-atom chains DST ≈ 50 mN/m, for ionic surfactants a steep increase of the accumulated foam is seen at DST < 70 mN/m followed by a more gradual increase of the initial foaming at DST < 60 mN/m.

**Figure 9.** Correlation between the foaming parameters and the surface coverage, $\Gamma(t_u)/\Gamma_{CMC}$ at $t_u = 10$ ms: (A) Volume of entrapped air after 10 shake cycles; (B) Volume of entrapped air after 100 shake cycles. The curves in (B) are guides to the eye. The other measures of the initial and maximum foaming, $V_{AMAX}/n_A$ and $V_{AMAX}$, show very similar trends as shown in Figures S5 in Supplementary materials.

**Figure 10.** Correlation between the foaming parameters and the dynamic Gibbs elasticity at universal surface age $t_u = 10$ ms: (A) Volume of entrapped air after 10 shake cycles; (B) Volume of entrapped air after 10 shake cycles. The other measures of initial and maximum foaming, $V_{AMAX}/n_A$ and $V_{AMAX}$, show very similar trends as seen in Figures S6 in Supplementary materials.

**Figure 11.** Correlation between the foaming parameters and the surface mobility factor, $h_S/h_{CR}$, at $t_u = 10$ ms, which accounts for the mobility of the film surfaces upon film thinning: (A) Volume of entrapped air after 10 shake cycles; (B) Volume of entrapped air after 100 shake cycles. The other measures of initial and maximum foaming, $V_{AMAX}/n_A$ and $V_{AMAX}$, show very similar trends as seen in Figures S7 in Supplementary materials.

**Figure 12.** Schematic presentation of the two main phenomena, surface mobility and foam film stabilization, in dynamic films during foaming: (A) For nonionic surfactants, very high surface coverage is needed to suppress the surface mobility, reduce the rate of film thinning (thus ensuring longer time for surfactant adsorption) and stabilize the film by steric forces. (B) For ionic surfactants, the faster adsorption and the strong electrostatic repulsion ensure film stabilization at much lower surface coverage. Note that the foam films could be very inhomogeneous in thickness and surfactant adsorption during foaming. Therefore, they could break locally, at some thinner regions with lower surfactant adsorption, even if the average film thickness and the average surfactant adsorption are relatively large.

**Figure 13.** (A) Picture of the apparatus used and (B,C) Foams generated after 10 shaking cycles from (B) 10 mM Brij 35 and (C) 10 mM SDS

# Notation

## Capital latin letters

$A_H$ – Hamaker constant

$C$ - concentration

    $C_S$ – total surfactant concentration

    $C_i$ - surfactant concentration of the *i*-th component in the solution

    $C_{EL}$ - concentration of the additional inorganic electrolyte.

$D$ – diffusion coefficient

    $D_{BC}$ - bulk diffusion coefficient of the surfactant molecules

    $D_{SC}$ - surface diffusion coefficient of the surfactant molecules

$E_G$ – Gibbs elasticity

$F$ - external force, pushing the bubble against a large interface, eq. 20

$I$ – total ionic strength

$K_B(\phi_1)$ - the dimensionless mobility function of the surfactant molecules, eq. 25

$P_C$ – capillary pressure

$R$ - universal gas constant

$R_F$ – radius of foam film

$R_b$ – bubble radius

$T$ - temperature

$V_A$ – volume of trapped air

    $V_{AMAX}$ - the maximum volume of the air which would be entrapped after a very large number of cycles, eq. 18

    $V_A(n)$ – volume of trapped air after *n* shaking cycles

$V$ – velocity of film thinning

    $V_{DR}$ - rate of film drainage in the presence of surfactants in the aqueous phase, eq. 19

    $V_{RE}$ - Reynolds velocity of thinning of planar film with tangentially immobile surfaces, eq. 20

**Small latin letters**

*a* - ionic activity

    $a_t$ – total ionic activity, eq. 9

    $a_S$ – surfactant ionic activity, eq. 9

$a_\sigma^2$ - characteristic time for surface tension decrease, eq. 13

*b* – parameter accounting for diffusion from the film interior, eq. 23

*h* – film thickness

    $h_{CR}$ – critical film thickness for film rupture, eq. 21

$h_S$ – parameter accounting for film surface mobility, eq. 24

$h_a$ - accounts for the surface activity of the surfactant, eq. 23

*n* – number of shake cycles

    $n_A$ - characteristic number of cycles at which $V_A$ reaches ≈ 63% of $V_{AMAX}$, eq. 18

$s_\sigma$ - parameter which accounts for the difference between the initial and the equilibrium surface tension, eq. 13

*t* – surface age

    $t_{age}$ - nominal surface age

    $t_u$ - universal surface age, eq. 12

$t_\Gamma$ - characteristic adsorption time, eq. 15θ

$z_0$ and $z_1$ - numerical coefficients in eq. 1 and eq. 10

**Capital greek letters**

Γ - adsorption

    $\Gamma_{tot}$ - the sum of all adsorbed species on the solution surface at given surfactant concentration

    $\Gamma_i$ – the adsorption of *i*-th component on the solution surface

    $\Gamma_\infty$ - maximal surfactant adsorption in dense adsorption layer

    $\Gamma_{CMC}$ – total surfactant adsorption at CMC

    $\Gamma(t)$ – dynamic total surfactant adsorption

    $\Gamma_{eq}$ - the equilibrium total surfactant adsorption at given surfactant concentration

    $\Gamma(0)$ - the initial adsorption at *t*=0, eq. 14

Φ - air volume fraction

**Small greek letters**

α - an average excluded area per molecule, which is equal to $1/\Gamma_\infty$

    $\alpha_{ii}$ - excluded area per molecule for i component

    $\alpha_{12}$ – defined by eq. 7

$\phi_1$ - surfactant volume fraction in the bulk, eq. 25

$\gamma_\pm$ - mean activity coefficient defined by eq. 8

$\eta_C$ - dynamic viscosity of surfactant solution

$\lambda^2$ - an apparatus constant for MBPM

π - the surface pressure

θ - surface coverage, eq. 17

    $\theta(t)$ – dynamic surface coverage defined as $\Gamma(t)/\Gamma_\infty$

    $\theta_{CMC}$ - surface coverage defined as $\Gamma(t)/\Gamma_{CMC}$

σ - surface tension

    $\sigma(t)$ - dynamic surface tension

    $\sigma_0$ - the surface tension of the aqueous phase without surfactant

    $\sigma(C_S)$ - the equilibrium surface tension at a certain surfactant concentration

    $\sigma_{CMC}$ – equilibrium surface tension at CMC

    $\sigma_{eq}$ - equilibrium surface tension

# Abbreviations

Brij 35 - polyoxyethylene-23 lauryl ether

Brij 58 - polyoxyethylene-20 cetyl ether

CMC – critical micellar concentration

CTAB - cetyltrimethylammonium bromide

DTAB - dodecyltrimethylammonium bromide

DST – dynamic surface tension

MBPM – maximum bubble pressure method

NaCl – sodium chloride

SDS – sodium dodecyl sulfate

Tween 20 - polyoxyethylenesorbitan monolaurate

Tween 40 - polyoxyethylene sorbitan monopalmitate

# Graphical abstract

"Foamability: Role of surfactant type and concentration"

(by Petkova et al.)

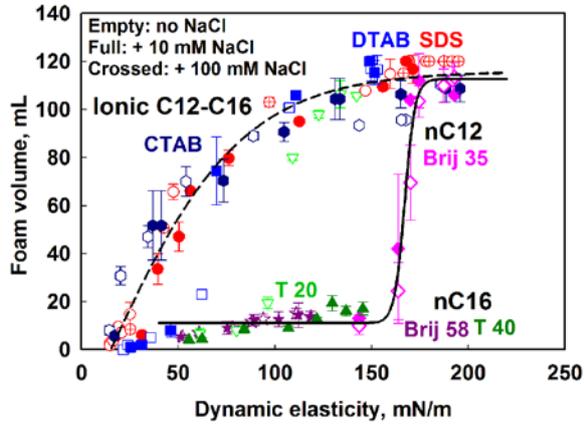
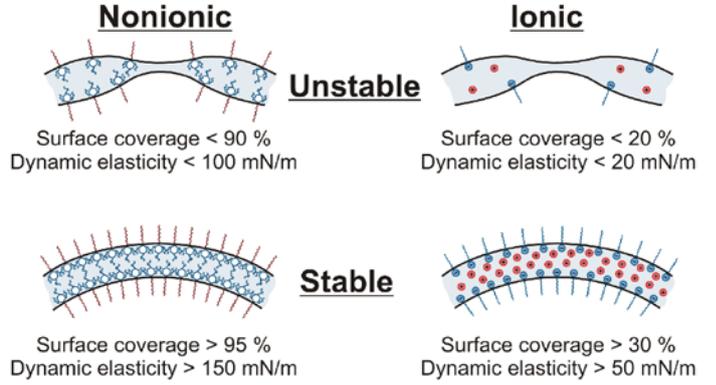

# Supplementary Information

# Foamability of aqueous solutions:
# Role of surfactant type and concentration

B. Damyanova,[1] S. Tcholakova,[1,*] M. Chenkova,[1] K. Golemanov,[1]
N. Denkov,[1] D. Thorley[2] and S. Stoyanov[3,4,5]

**Section S1.** Additional figures.

**Figure S1.** (A,B) Dynamic surface tension at 10 ms; (C,D) Surfactant adsorption at 10 ms; (E, G) Gibbs elasticity at 10 ms; (G,H) Surface coverage, $\Gamma/\Gamma_{CMC}$, vs. C/CMC for nonionic surfactants (A,C,E,G) and ionic surfactants (B,D,F,H) without electrolyte (empty symbols), with 10 mM NaCl (full symbols) and with 100 mM NaCl (crossed symbols).

**Figure S2.** Maximum volume of trapped air for (A) nonionic and (B) ionic surfactants, as function of C/CMC.

**Figure S3.** (A) Initial rate of air entrapment, and (B) maximum volume of trapped air as functions of dynamic surface tension determined at $t_u = 10$ ms.

**Figure S4.** Correlation between the foaming parameters and the dynamic surface tension at $t_u = 2$ ms: (A) Volume of entrapped air after 10 shake cycles; (B) Volume of entrapped air after 100 shake cycles.

**Figure S5.** (A) Initial rate of air entrapment, and (B) maximum volume of trapped air as functions of surface coverage, $\Gamma(t_u)/\Gamma_{CMC}$ at $t_u = 10$ ms.

**Figure S6.** Correlation between the foaming parameters and the dynamic Gibbs elasticity at universal surface age $t_u = 10$ ms: (A) Initial rate of air entrapment; (B) maximum volume of trapped air.

**Figure S7.** Correlation between the foaming parameters and the surface mobility factor, $h_S/h_{CR}$, at $t_u = 10$ ms, which accounts for the mobility of the film surfaces upon film thinning: (A) Initial rate of air entrapment; (B) maximum volume of trapped air.

**Section S2.** Approximate expressions for the parameters characterizing the mobility of surfactant molecules in the bulk solutions and on the foam film surface.

**Section S1. Additional figures.**

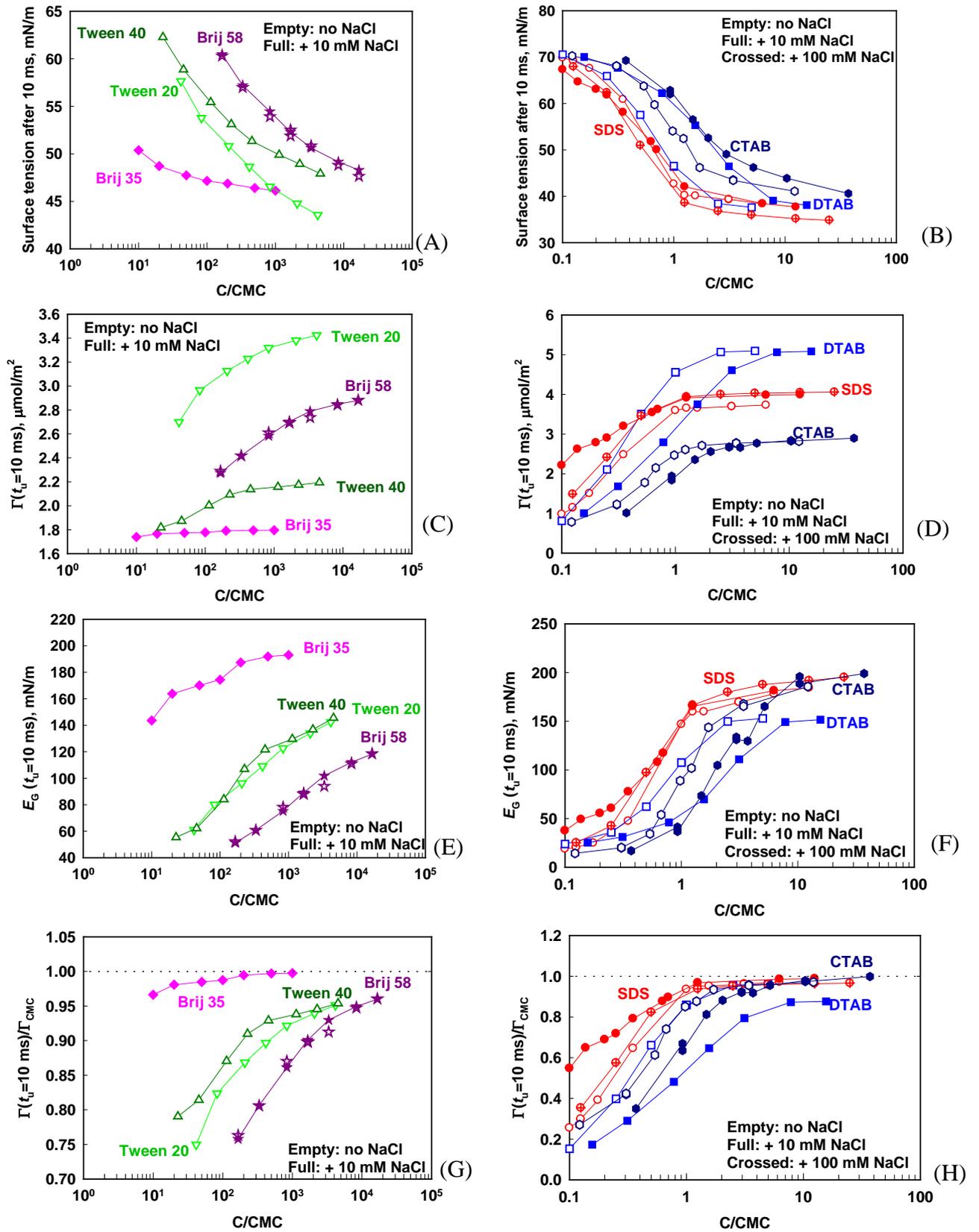

**Figure S1.** (A,B) Dynamic surface tension at 10 ms; (C,D) Surfactant adsorption at 10 ms; (E, G) Gibbs elasticity at 10 ms; (G,H) Surface coverage, $\Gamma/\Gamma_{CMC}$, vs. C/CMC for nonionic surfactants (A,C,E,G) and ionic surfactants (B,D,F,H) without electrolyte (empty symbols), with 10 mM NaCl (full symbols) and with 100 mM NaCl (crossed symbols).

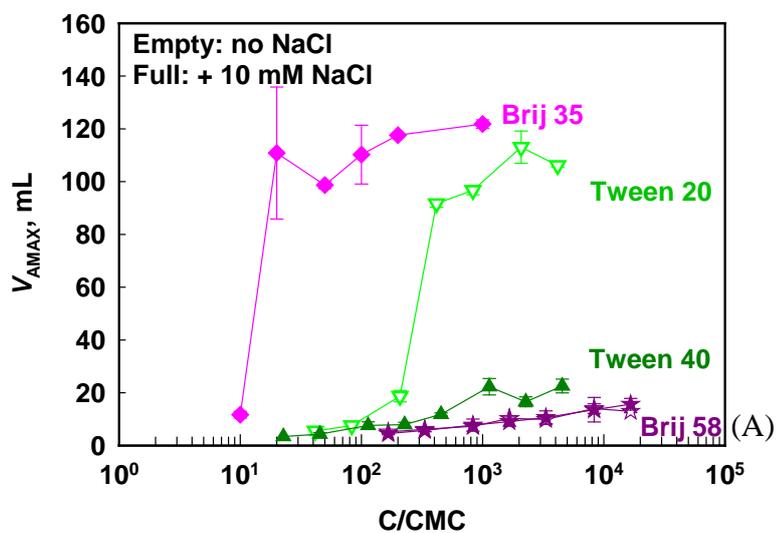

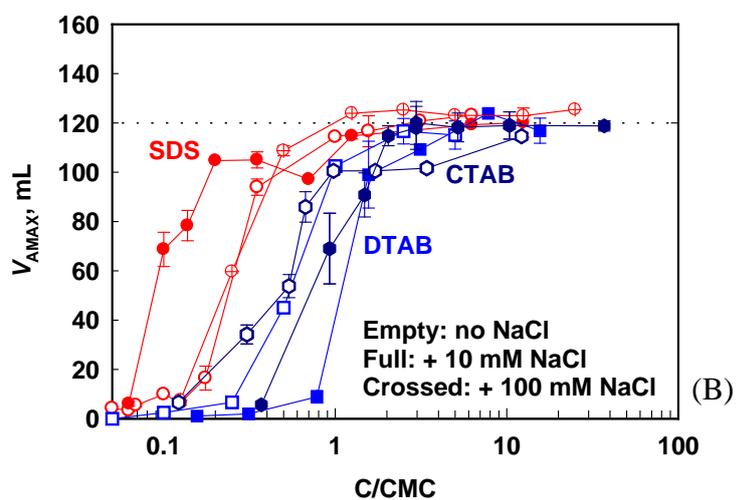

**Figure S2.** Maximum volume of trapped air for (A) nonionic and (B) ionic surfactants, as function of C/CMC.

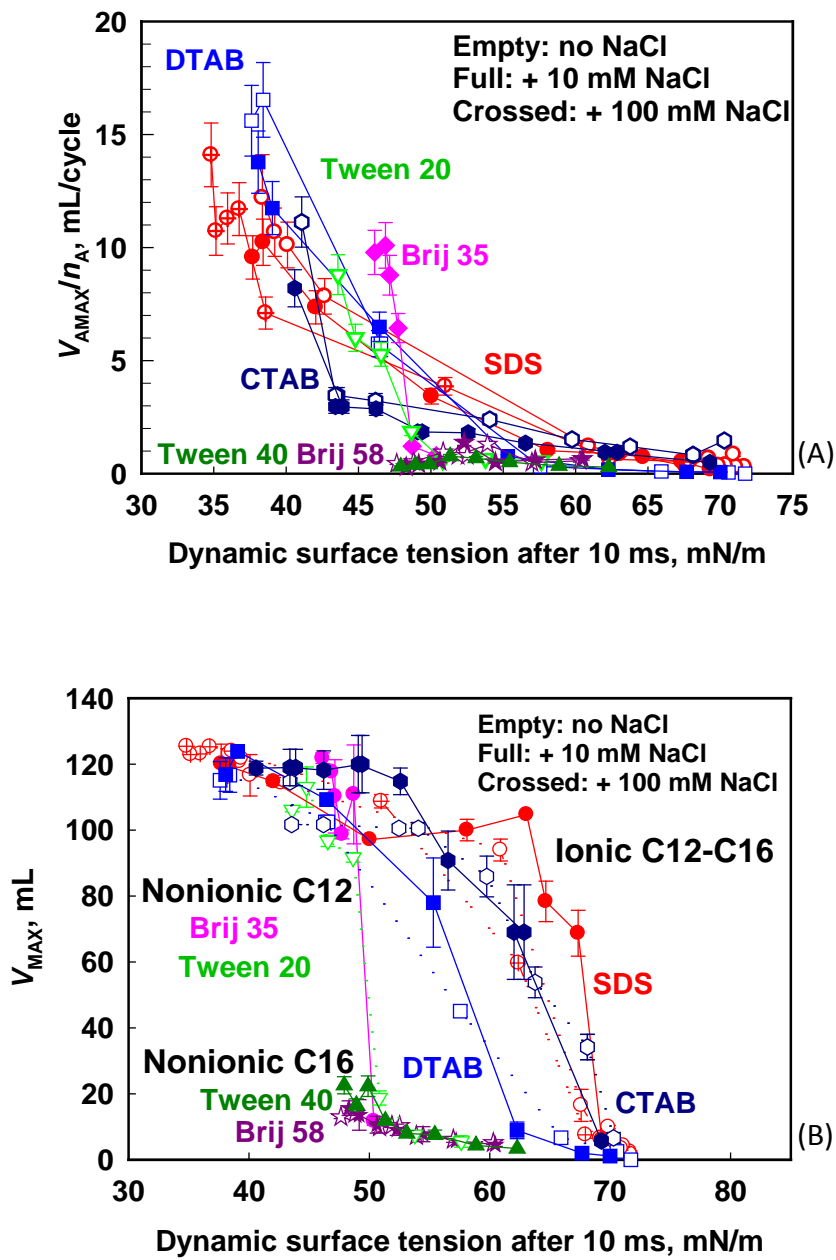

**Figure S3.** (A) Initial rate of air entrapment, and (B) maximum volume of trapped air as functions of dynamic surface tension determined at $t_u = 10$ ms.

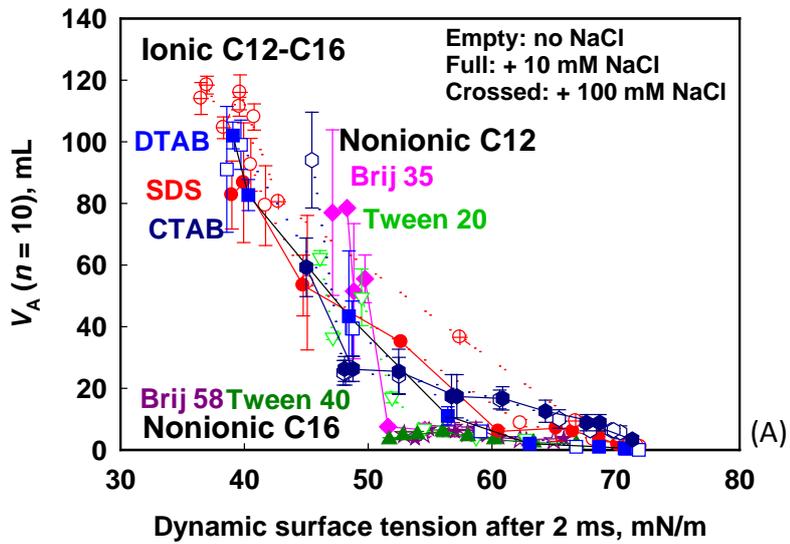
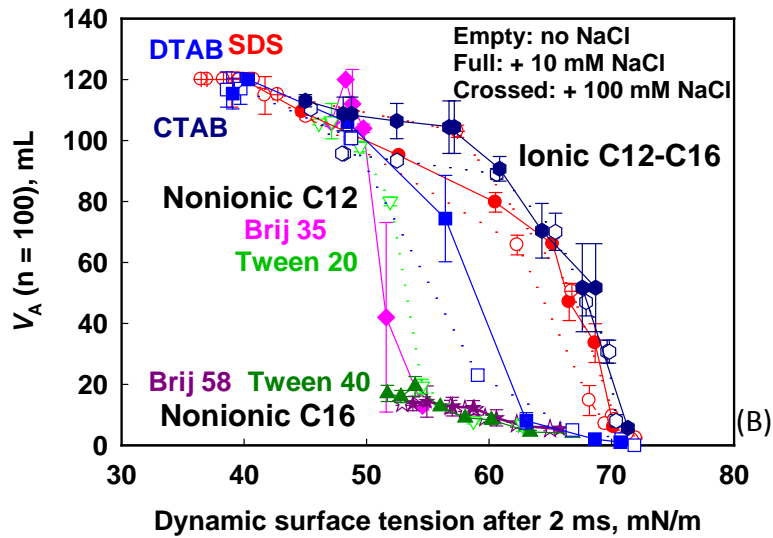

**Figure S4.** Correlation between the foaming parameters and the dynamic surface tension at $t_u$ = 2 ms: (A) Volume of entrapped air after 10 shake cycles; (B) Volume of entrapped air after 100 shake cycles.

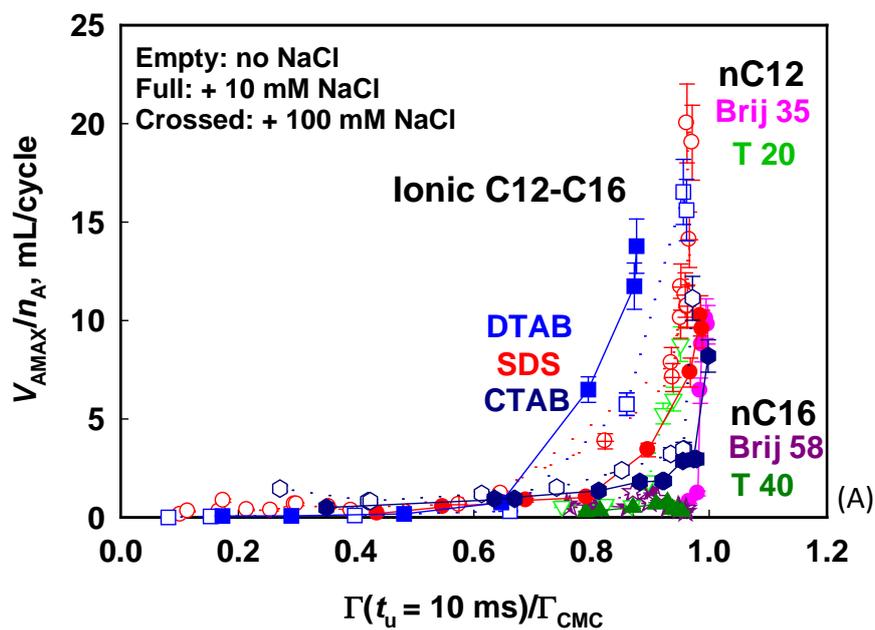

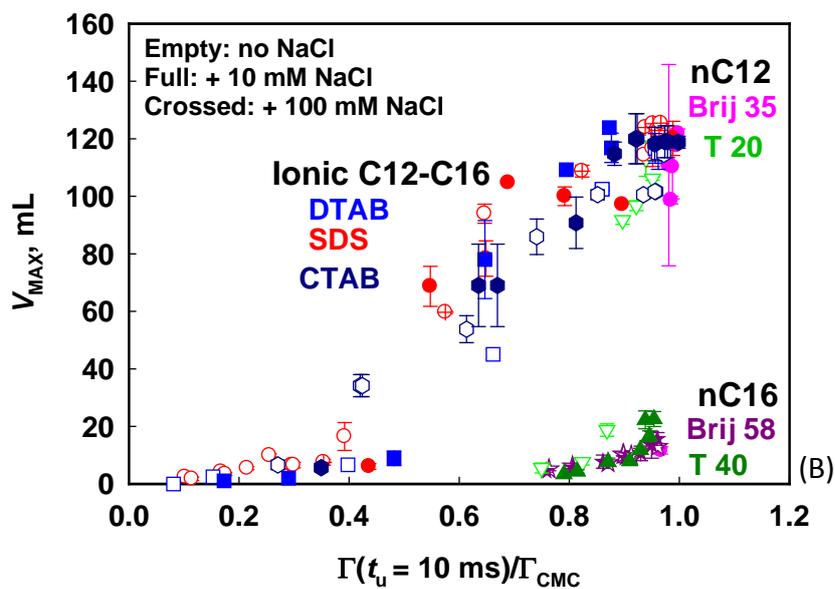

**Figure S5.** (A) Initial rate of air entrapment, and (B) maximum volume of trapped air as functions of surface coverage, $\Gamma(t_u)/\Gamma_{CMC}$ at $t_u = 10$ ms.

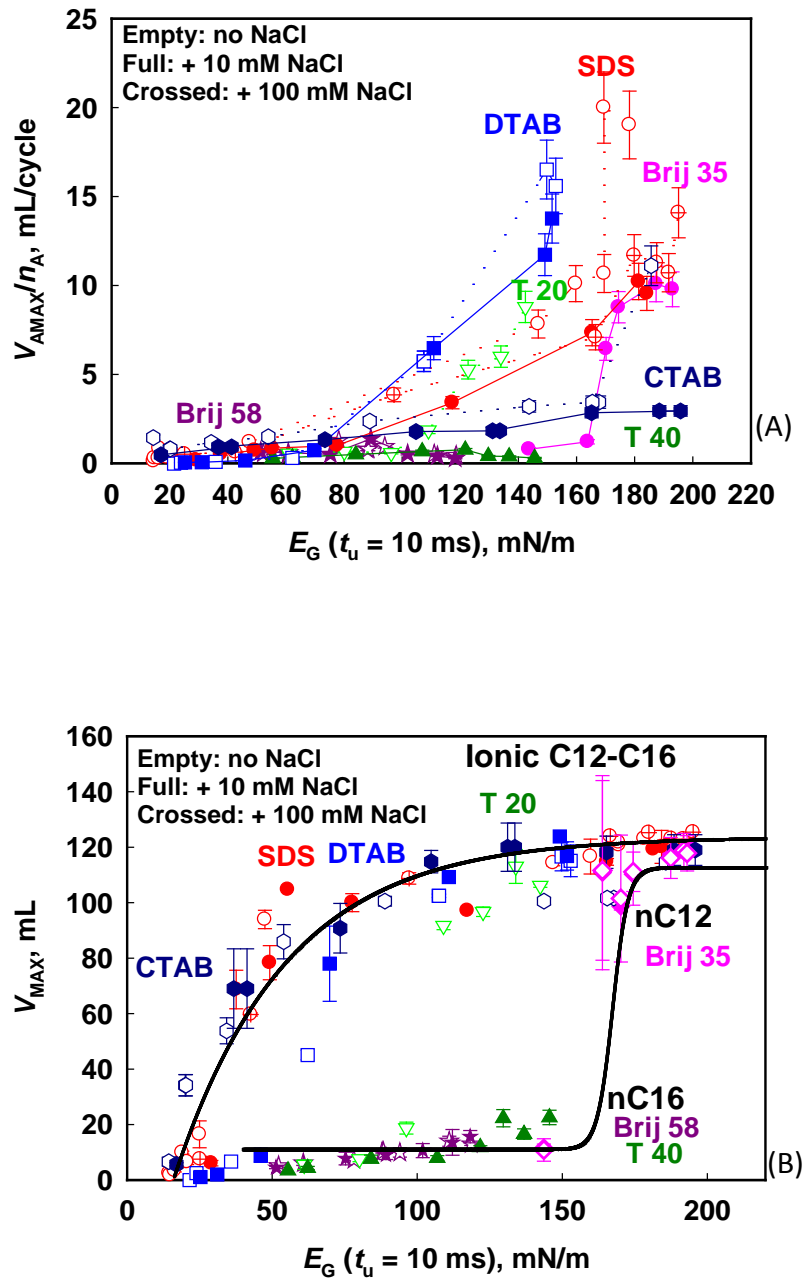

**Figure S6.** Correlation between the foaming parameters and the dynamic Gibbs elasticity at universal surface age $t_u = 10$ ms: (A) Initial rate of air entrapment; (B) maximum volume of trapped air.

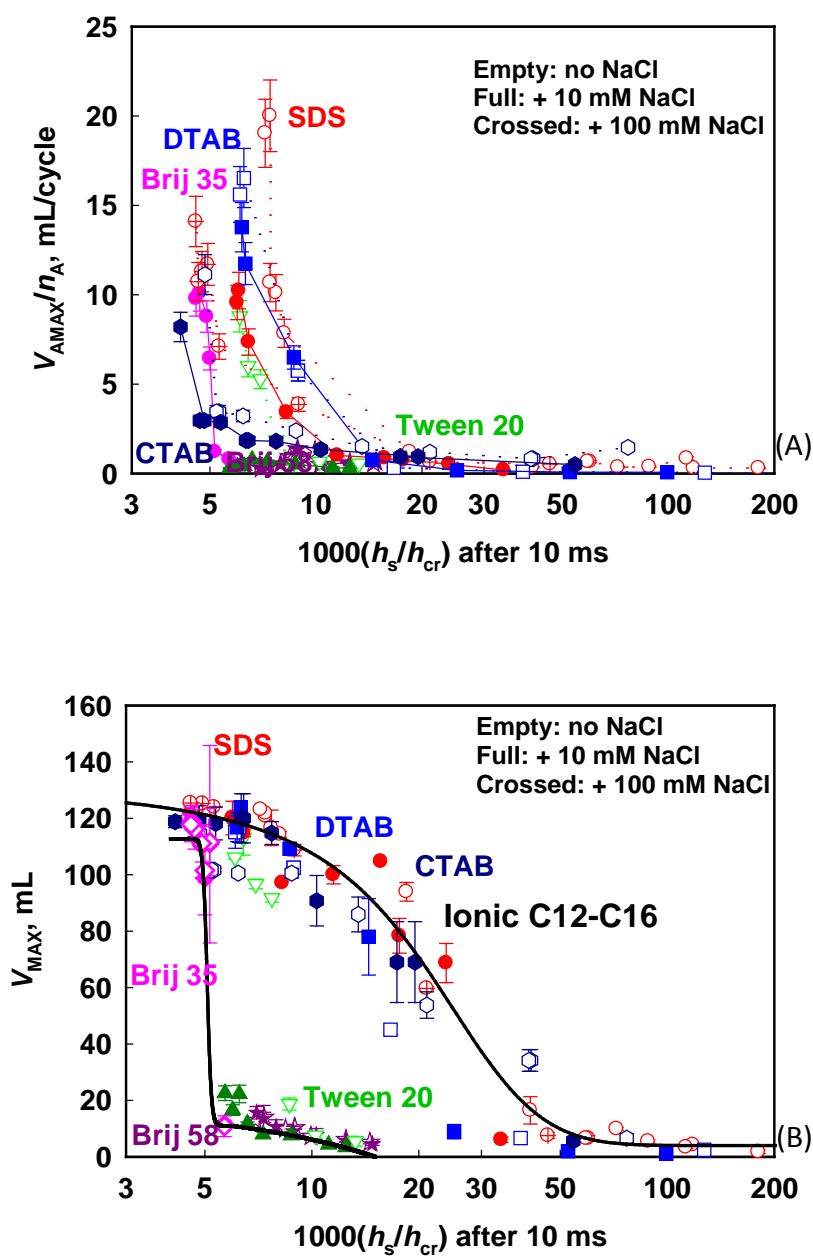

**Figure S7.** Correlation between the foaming parameters and the surface mobility factor, $h_S/h_{CR}$, at $t_u = 10$ ms, which accounts for the mobility of the film surfaces upon film thinning: (A) Initial rate of air entrapment; (B) maximum volume of trapped air.

## Section S2. Approximate expressions for the parameters characterizing the mobility of surfactant molecules in the bulk solutions and on the foam film surface.

As explained in Section 4.5 of the main text, $K_S(\Gamma_1)$ is the dimensionless mobility function in the adsorption layer [S1] which depends on the hydrodynamic interactions with the other surfactant molecules in this layer. Typically, this mobility function is smaller than unity, $K_S < 1$, because the hydrodynamic interactions slow down the molecule diffusion due to the increased hydrodynamic friction. In contrast, the Gibbs elasticity, $E_G$, appearing in Eq. (25), accelerates the surface diffusion as it acts as a thermodynamic force, pushing the surfactant molecules to move along the gradient of the surfactant chemical potential (viz. along the gradient of adsorption). Thus, $D_{SC}$ may differ significantly from $D_{S0}$ and eq. (25) should be used to account properly for the surface diffusion of surfactant. Using Volmer adsorption isotherm, eq. (5), we may substitute $E_G$ from eq. (16) and $D_{SC}$ from eq. (25) into eqs. (23)-(24) to obtain:

$$b = \frac{3\eta C}{\beta_{B0}\Gamma^2} \qquad h_S = \frac{6\eta K_S(\Gamma)}{\beta_{S0}\Gamma} \tag{S1}$$

where $C$ is the monomer surfactant concentration. For solutions with concentration lower than the CMC, this is the actual surfactant concentration. For the solutions above the CMC, $C$ is equal to CMC. Note that the alternative assumption that $C$ is always equal to the total surfactant concentration, both below and above the CMC, gave predictions which deviated qualitatively from the observed experimental trends.

The surface and bulk friction coefficients of the surfactant molecules, $\beta_{S0}$ and $\beta_{B0}$, can be determined from the molecular dimensions of the studied surfactant molecules. For $\beta_{B0}$ we assume that the surfactant molecules can be described as prolate ellipsoids with semi-axes $r$, $r$ and $pr$ and the respective friction coefficient is presented as [S2]:

$$\beta_{B0} = 6\pi\eta r \frac{\text{Arccosh } p}{\sqrt{p^2-1}} \tag{S2}$$

Here $r$ is the radius of the hydrophilic group and $p$ is the ratio between the length of the hydrophobic tail and the radius of the hydrophilic head group. To determine the length of the hydrophobic tail we used the expression [S3]:

$$l = 0.154 + 0.1265n \quad \text{in nm} \tag{S3}$$

where $n$ is the number of C-atom in the surfactant tail. To determine the value of the surface friction coefficient, $\beta_{S0}$, we assumed that the surfactant molecules move along the interface as thin discs of radius $r$ [S4]:

$$\beta_{S0} = 12\eta\, r \tag{S4}$$

The explicit dependence of $K_S(\Gamma)$ is unknown but for approximate estimates one could use a 2D approximation of the known expressions for bulk particle suspensions, $K_S \approx (1-\theta)$, to account for the reduced mobility in the presence of the other adsorbed molecules. This simple expression accounts correctly for the two limiting cases: $K_S \to 1$ for diluted adsorption layers and $K_S \to 0$ for densely packed adsorption layers, without introducing other unknown constants. Substituting the expression $K_S \approx (1-\theta)$ and the expressions for $\beta_{S0}$ and $\beta_{B0}$ from eqs. (S4) and (S2), respectively, we find the following approximate expressions for $b$ and $h_s$ which are used for numerical calculations and comparison with the experimental data:

$$b = \frac{C}{2\pi r \dfrac{\operatorname{Arccosh} p}{\sqrt{p^2-1}} \Gamma^2} \qquad h_S = \frac{(1-\theta)}{2r\Gamma} \tag{S5}$$